\documentclass[amsmath,nofootinbib,notitlepage,prd,superscriptaddress,twocolumn]{revtex4-2}

\usepackage[T1]{fontenc}
\usepackage[usenames,dvipsnames]{xcolor}
\usepackage{aas_macros,graphicx,hyperref,orcidlink}
\hypersetup{colorlinks=true,citecolor=RoyalBlue,linkcolor=RoyalBlue,urlcolor=RoyalBlue}

\newcommand{\ab}[1]{\left|#1\right|}
\newcommand{\av}[1]{\left\langle#1\right\rangle}
\newcommand{\br}[1]{\left[#1\right]}
\newcommand{\cu}[1]{\left\{#1\right\}}
\newcommand{\pa}[1]{\left(#1\right)}
\newcommand{\ed}{\mathop{}\!\mathrm{d}}
\DeclareMathOperator\arcsinh{arcsinh}

\begin{document}

\title{Prediction for the interferometric shape of the first black hole photon ring\texorpdfstring{\vspace{-4pt}}{}}

\author{Alejandro C\'ardenas-Avenda\~no\,\orcidlink{0000-0001-9528-1826}} 
\affiliation{Princeton Gravity Initiative, Princeton University, Princeton, New Jersey 08544, USA}

\author{Alexandru Lupsasca\,\orcidlink{0000-0002-1559-6965}}
\affiliation{Department of Physics \& Astronomy, Vanderbilt University, Nashville, Tennessee 37212, USA}

\begin{abstract}

Black hole images are theoretically predicted---under mild astrophysical assumptions---to display a stack of lensed ``photon rings'' that carry information about the underlying spacetime geometry.
Despite vigorous efforts, no such ring has been observationally resolved thus far.
However, planning is now actively under way for space missions targeting the first (and possibly the second) photon rings of the supermassive black holes M87* and Sgr A*.
In this work, we study interferometric photon ring signatures in time-averaged images of Kerr black holes surrounded by different astrophysical profiles.
We focus on the first, most easily accessible photon ring, which has a larger width-to-diameter ratio than subsequent rings and whose image consequently lacks a sharply defined diameter.
Nonetheless, we show that it does admit a precise angle-dependent diameter \textit{in visibility space}, for which the Kerr metric predicts a specific functional form that tracks the critical curve.
We find that a measurement of this interferometric ring diameter is possible for most astrophysical profiles, paving the way for precision tests of strong-field general relativity via near-future observations of the first photon ring.
\end{abstract}

\maketitle

\section{Introduction}

Theoretical work \cite{Gralla2019,JohnsonLupsasca2020,GrallaLupsasca2020a,Chael2021} predicts that---under some mild assumptions---images of an astrophysical Kerr black hole generically display a stack of nested ``photon rings,'' each of which is a strongly lensed image of the main emission superimposed on top of the direct image.
These rings may be labeled by the number $n$ of half-orbits executed around the black hole by their constitutive photons on their way from source to observer.
The full set of $n\geq1$ rings is often collectively referred to as ``\textit{the} photon ring'': a striking feature that dominates simulated black hole images, and a signature stamp of strong gravity (Fig.~\ref{fig:RingDiameters}).

Despite vigorous efforts \cite{EHT2019e,EHT2019f,Lockhart2021,Broderick2022}, Event Horizon Telescope (EHT) observations from Earth of the supermassive black holes M87* and Sgr\,A* \cite{EHT2019a,EHT2022a} have yet to experimentally resolve any photon ring \cite{Lockhart2022,Tiede2022}.
Space missions targeting their first ($n=1$)---and possibly second ($n=2$)---photon rings are now being planned \cite{Gurvits2022,Kurczynski2022,GLM2020}.
While a theoretical prediction for the interferometric signature of the $n\ge2$ rings has already been derived \cite{Gralla2020,GrallaLupsasca2020c,GLM2020,Staelens2023} and explored \cite{Paugnat2022,Vincent2022,CardenasAvendano2022}, a sharp prediction for the interferometric signature of the more readily accessible $n=1$ ring is still lacking.
This paper formulates such a prediction (Sec.~\ref{subsec:Summary}).

\section{Photon ring images}
\label{sec:RingImages}

The lensing behavior of the Kerr geometry confers two properties to the appearance of the photon ring.
First, since each subring is a mirror image of its predecessor, the full photon ring must exhibit a self-similar substructure, which in the limit $n\to\infty$ is completely characterized by three critical exponents $\gamma$, $\delta$, and $\tau$ that respectively control the demagnification, rotation, and time delay of successive images \cite{GrallaLupsasca2020a}.
The analytically known parameters $(\gamma,\delta,\tau)$ depend only on the mass and spin of the black hole---as well as the photon orbital radius \cite{JohnsonLupsasca2020,GrallaLupsasca2020a}---and may in principle be measured from observations of light echoes or their characteristic pattern of autocorrelations \cite{Hadar2021,GrallaLupsasca2023}.
Since successive subrings are exponentially demagnified by $\sim e^{-\gamma}$, the large-$n$ rings quickly become so narrow in the image plane of a distant observer that they may---to a very good approximation---be regarded as infinitely thin, mathematical curves $\mathcal{C}_n$.
In fact, this is an excellent approximation for $n\ge2$, as the second ($n=2$) photon ring already appears extremely thin; typically, only the first ($n=1$) ring displays a noticeable thickness (Fig.~\ref{fig:RingDiameters}).

The second property is closely tied to the exponential subring demagnification: the photon rings must converge (exponentially fast in $n$) to a theoretical ``critical curve'' in the image plane of an observer, which corresponds to the image of the black hole's (asymptotically) bound photon orbits.
First derived by Bardeen \cite{Bardeen1973}, this analytically known curve---call it $\tilde{\mathcal{C}}$---delineates the apparent cross-section of a black hole in the sky. It is fully determined by the Kerr geometry (together with the observer inclination  $\theta_{\rm o}$).
Thus, the critical curve is the ``$n\to\infty$ photon ring''
\begin{align}
	\label{eq:Convergence}
    \tilde{\mathcal{C}}=\mathcal{C}_\infty
    \equiv\lim_{n\to\infty}\mathcal{C}_n,
\end{align}
and indeed, photons that appear exactly on $\tilde{\mathcal{C}}$ lie on null rays that were unstably trapped (in the far past) within a region of spacetime just outside the event horizon, which is now known as the ``photon shell'' \cite{JohnsonLupsasca2020,GrallaLupsasca2020a}; see also \cite{Teo2003, Perlick2004,Grenzebach2014,Teo2021,Hadar2022}.

The preceding discussion leads to a simple description of the large-$n$ subring images: they appear as thin curves $\mathcal{C}_n$ that closely track the critical curve $\tilde{\mathcal{C}}=\mathcal{C}_\infty$, with the deviations exponentially suppressed in $n$.
As illustrated in the bottom-right panel of Fig.~\ref{fig:RingDiameters}, this description is already valid for $n=2$: the second photon ring looks like a bright, narrow curve that sits exactly atop $\tilde{\mathcal{C}}$ (drawn as a dashed black line).
On the other hand, the bottom-left panel of Fig.~\ref{fig:RingDiameters} shows why this description fails for $n=1$: the first ring has a significant width, and its shape visibly deviates from that of the (dashed black) critical curve.

Moreover, the appearance of the $n=1$ photon ring---both its thickness and deviation from $\tilde{\mathcal{C}}$---can significantly vary with the choice of astrophysical source.
This leads to the central question of this paper: can one produce a sharp theoretical prediction for the $n=1$ ring shape?

\section{Interferometric ring signatures}
\label{sec:RingSignatures}

The key to predicting the $n=1$ ring shape is to work in Fourier space. 
Interferometers sample the radio visibility
\begin{align}
	\label{eq:ComplexVisibility}
	V(\mathbf{u})=\int I_{\rm o}(\mathbf{x}_{\rm o})e^{-2\pi i\mathbf{u}\cdot\mathbf{x}_{\rm o}}\ed^2\mathbf{x}_{\rm o},
\end{align}
which is the Fourier transform of the sky image $I_{\rm o}(\mathbf{x}_{\rm o})$.
The dimensionless baseline $\mathbf{u}$ sampled by two elements is the distance separating them in the plane perpendicular to the line of sight, in units of the observation wavelength.

An image of an infinitely thin ring produces a visibility with a characteristic ringing pattern, whose periodicity at polar angle $\varphi$ in the baseline plane $\mathbf{u}=(u,\varphi)$ is set by the (precise, well-defined) diameter of the ring at the corresponding angle $\phi=\varphi$ in the image plane $\mathbf{x}_{\rm o}=(\rho,\phi)$.
On the other hand, if the ring has some thickness, then its image lacks a well-defined diameter, but nevertheless its corresponding visibility still rings with a characteristic periodicity, from which a sharp notion of angle-dependent ``interferometric ring diameter'' $d_\varphi$ can thus be derived.

The main idea of this paper is to \textit{define} the diameter $d_\varphi^{(1)}$ of the first ($n=1$) photon ring from the periodicity of its ringing interferometric signature.
In the remainder of this section, we will describe precisely how $d_\varphi^{(1)}$ may be recovered from the visibility \eqref{eq:ComplexVisibility} that is directly probed by an interferometer, and formulate a guess for its functional form.
In the rest of the paper, we will then study a set of astrophysical source models around a Kerr black hole and show that the angle-dependent diameter $d_\varphi^{(1)}$ of their first photon ring follows this functional form to high accuracy.

\subsection{Interferometric signature of a zero-width ring}

To make sense of the preceding remarks, the first step is to consider perfectly thin rings, or more generally, images that consist of an infinitely narrow, bright curve $\mathcal{C}$.
If $\mathcal{C}$ is closed and convex,\footnote{If $\mathcal{C}$ is not closed and convex, then it does not admit a single normal-angle parametrization and must be covered by multiple segments $(x_i(\varphi),y_i(\varphi))$ \cite{GrallaLupsasca2020c}; we will not consider such cases here.} then its shape can always be parametrized in the Cartesian image plane $\mathbf{x}_{\rm o}=(\alpha,\beta)$ by the normal angle angle $\varphi$ to the curve \cite{GrallaLupsasca2020c},
\begin{align}
\label{eq:NormalParameterization}
	\mathcal{C}=\cu{(\alpha(\varphi),\beta(\varphi))\,|\,\varphi\in[0,2\pi)}.
\end{align}
In practice, given another parametrization $(\alpha(\sigma),\beta(\sigma))$ of $\mathcal{C}$, this parameterization may be obtained by solving
\begin{align}
	\tan{\varphi(\sigma)}=-\frac{\alpha'(\sigma)}{\beta'(\sigma)}
\end{align}
for the normal angle $\varphi(\sigma)$ along the curve, and then plugging the inverse $\sigma(\varphi)$ into the original parametrization.
Thereafter, one can compute the \textit{projected position} of $\mathcal{C}$,
\begin{align}
	\label{eq:ProjectedPosition}
	f(\varphi)\equiv x(\varphi)\cos{\varphi}+y(\varphi)\sin{\varphi}.
\end{align}
This function completely encodes the shape of $\mathcal{C}$, which may still be recovered via the inverse relations
\begin{subequations}
\begin{align}
	x(\varphi)&=f(\varphi)\cos{\varphi}-f'(\varphi)\sin{\varphi},\\
	y(\varphi)&=f(\varphi)\sin{\varphi}+f'(\varphi)\cos{\varphi}.
\end{align}
\end{subequations}
From an interferometric perspective, however, it is most natural to describe $\mathcal{C}$ via its projected position \eqref{eq:ProjectedPosition}, which turns out to be most closely connected to the visibility \eqref{eq:ComplexVisibility} of $\mathcal{C}$ that an interferometer would directly sample.

To connect $f(\varphi)$ to interferometric observables, we first decompose it into its parity-even and parity-odd parts,
\begin{subequations}
\label{eq:ParityDecomposition}
\begin{align}
	d_\varphi&\equiv f(\varphi)+f(\varphi+\pi),\\
	C_\varphi&\equiv\frac{1}{2}\br{f(\varphi)-f(\varphi+\pi)},
\end{align}
\end{subequations}
which are the angle-dependent projected diameter and projected centroid displacement at angle $\varphi$ in the image of $\mathcal{C}$, respectively---see \cite{GrallaLupsasca2020c} for further discussion of their geometric interpretation.
Here, we simply note that $d_\varphi$ and $C_\varphi$ carry all the information about the shape of $\mathcal{C}$ that was stored in the projected position function, since
\begin{align}
	f(\varphi)=\frac{d_\varphi}{2}+C_\varphi.
\end{align}
While it may seem that we have now doubled the degrees of freedom needed to describe $\mathcal{C}$, that is not in fact the case because, as defined in \eqref{eq:ParityDecomposition}, $d_\varphi$ and $C_\varphi$ only range over $[0,\pi)$, repeating periodically thereafter.
Geometrically, this makes sense since the diameter and centroid are only defined for pairs of points $(\varphi,\varphi+\pi)$ around the curve.\footnote{We also note that the $\pi$-periodicity of $d_\varphi$, $C_\varphi$, and $\alpha_\varphi^{\rm L,R}$ ensures that the Fourier transform \eqref{eq:UniversalVisibility} satisfies $V(u,\varphi+\pi)=V^*(u,\varphi)$, as required by its definition \eqref{eq:ComplexVisibility} for a real image $I_{\rm o}(\mathbf{x}_{\rm o})=I_{\rm o}^*(\mathbf{x}_{\rm o})$.}

\begin{figure*}[ht!]
    \centering
    \includegraphics[width=\textwidth]{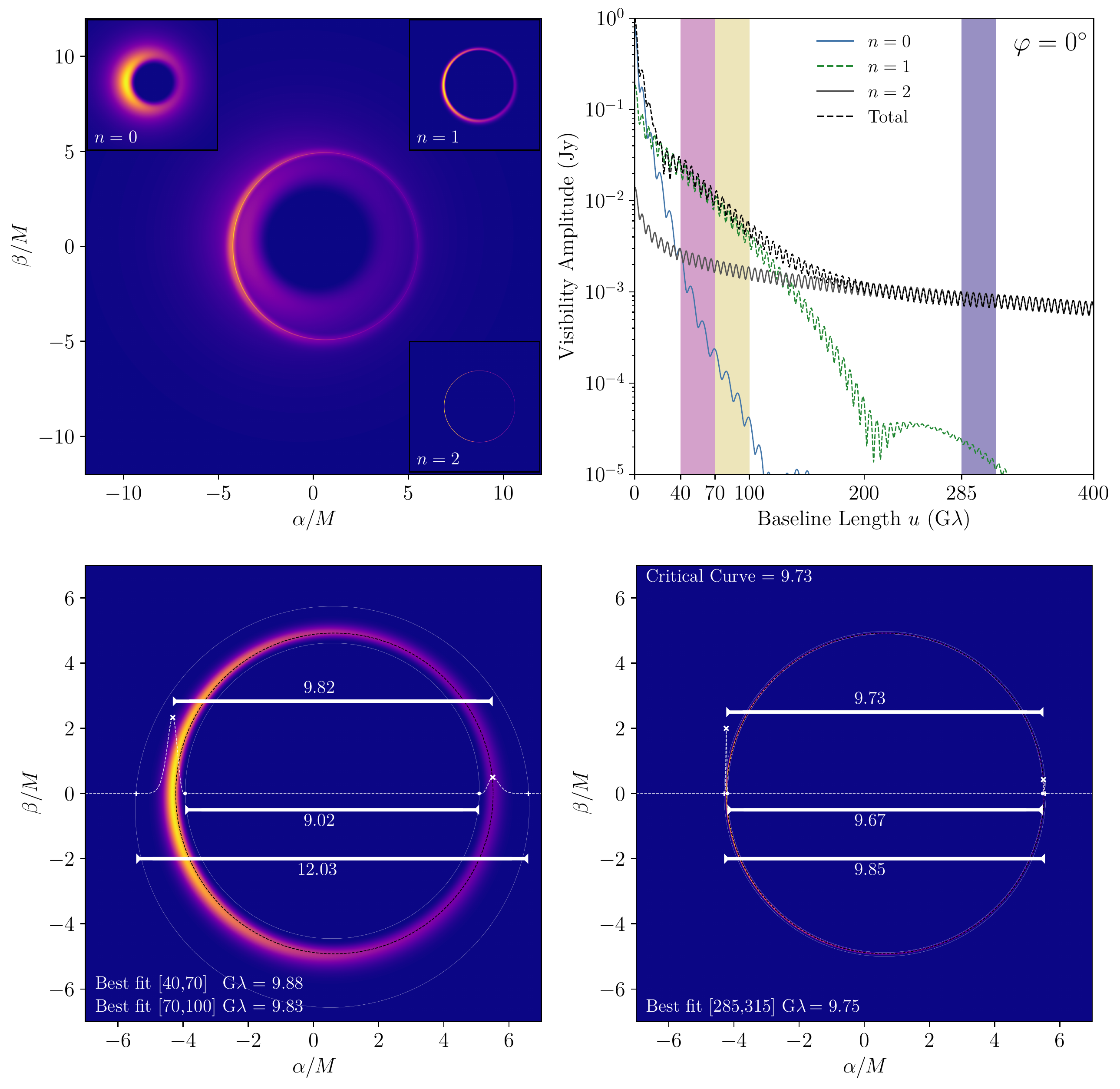}
    \caption{\textbf{Top left:} Adaptively ray-traced (with \texttt{AART} \cite{CardenasAvendano2022}) image of a stationary, axisymmetric, equatorial source with a radial profile given by \eqref{eq:RadialProfile} with $\mu=3r_+/2$, $\gamma=0$, and $\vartheta=M$.
    The inset panels decompose the image into its photon-orbit layers: the direct ($n=0$) image and the first two ($n=1$ and $n=2$) photon rings.
    \textbf{Top right:} The corresponding visibility amplitudes for a spin-perpendicular ($\varphi=0^\circ$) cut across the total image (black dashed line) and across each image layer.
    \textbf{Bottom left:} The image of the $n=1$ photon ring only.
    Three characteristic diameters are measured along a horizontal cut of the intensity profile, corresponding to (from top to bottom): the distance between the location of the peaks in the intensity ($9.82M$), the distance between the inner edges of the ``$n=1$ lensing band \cite{CardenasAvendano2022}'' ($9.02M$), and the distance between the outer edges of the band ($12.03M$).
    The diameter $d_{0^\circ}^{(1)}$ inferred from the characteristic ringing of the total visibility amplitude in two baseline windows $[40,70]\,$G$\lambda$ and $[70,100]\,$G$\lambda$ is reported.
    \textbf{Bottom right:} Same as in the bottom-left panel, but for the $n=2$ ring and with a projected diameter $d_{0^\circ}^{(2)}$ inferred from the ringing in the baseline window $[285,315]\,$G$\lambda$.
    Here, the black hole spin is $a/M=94\%$, the observer inclination is $\theta_{\rm o}=17^\circ$, and the critical curve has a diameter of $9.73M$ along the considered horizontal cut.}
    \label{fig:RingDiameters}
\end{figure*}

We now come to the key conclusion of \cite{Gralla2020}: the Fourier transform of an infinitely narrow curve $\mathcal{C}$ with projected diameter $d_\varphi$ and projected centroid $C_\varphi$ is approximately
\begin{align}
	\label{eq:UniversalVisibility}
	V(\mathbf{u})\approx\frac{e^{-2\pi iC_\varphi u}}{\sqrt{u}}\br{\alpha_\varphi^{\rm L}e^{-\frac{i\pi}{4}+i\pi d_\varphi u}+\alpha_\varphi^{\rm R}e^{\frac{i\pi}{4}-i\pi d_\varphi u}},
\end{align}
where the coefficients $\alpha_\varphi^{\rm L,R}=\alpha_{\varphi+\pi}^{\rm R,L}>0$ encode the polar intensity profile around the curve, and the approximation holds for $ud_\varphi\gg1$.
In particular, the visibility amplitude is a damped oscillation with radial periodicity $\Delta u=1/d_\varphi$ inside an envelope with a weak $\sqrt{u}$ power-law falloff,
\begin{align}
	\label{eq:UniversalAmplitude}
	\ab{V(\mathbf{u})}&\approx\sqrt{\frac{\pa{\alpha_\varphi^{\rm L}}^2+\pa{\alpha_\varphi^{\rm R}}^2+2\alpha_\varphi^{\rm L}\alpha_\varphi^{\rm R}\sin\pa{2\pi d_\varphi u}}{u}},
\end{align}
which depends only on the projected diameter $d_\varphi$.\newpage

On the other hand, the projected centroid $C_\varphi$ is only encoded in the visibility phase, which we will henceforth ignore as it is significantly harder to measure, and beyond the reach of presently envisioned $n=1$ ring observations.

\subsection{Interferometric signature of the photon ring}

So far, we have argued that an image-plane curve with angle-dependent diameter $d_\varphi$ produces an interferometric response on long baselines $u\gg1/d_\varphi$ that is completely captured by the visibility \eqref{eq:UniversalVisibility}.
In particular, its visibility amplitude displays a characteristic ringing signature \eqref{eq:UniversalAmplitude} whose periodicity at angle $\varphi$ in the baseline plane encodes the image diameter $d_\varphi$ of the curve at image angle $\phi=\varphi$.

Strictly speaking, this discussion only pertains to zero-width curves.
Intuitively, however, if the curve were in fact a very narrow ring with a small width-to-diameter ratio $w/d\ll1$, then we would expect it to produce the same response in an interferometer limited to sampling only baselines $uw\ll1$ too short to resolve the ring width.
This intuition was in fact proved in \cite{Gralla2020}, which computed the Fourier transform of such a thin ring to leading order in $w/d\ll1$, and found that the same approximation \eqref{eq:UniversalVisibility} to the complex visibility still holds in the baseline range
\begin{align}
	\label{eq:UniversalRegime}
    \frac{1}{d}\ll u\ll\frac{1}{w}.
\end{align}
This range is aptly called the ``universal regime'' since all thin rings produce the same universal signature \eqref{eq:UniversalVisibility}--\eqref{eq:UniversalAmplitude} on these baselines, regardless of their radial profile: it is only on even longer baselines $u\gtrsim1/w$ that a ring profile can be resolved and different rings can be distinguished.

For a ring with a smooth radial profile, the visibility ought to decay very rapidly once its width is resolved.
Therefore, as first noted in \cite{JohnsonLupsasca2020} and extensively reviewed in \cite{GrallaLupsasca2020a,Paugnat2022,CardenasAvendano2022}, the sequence of exponentially demagnified photon rings described in Sec.~\ref{sec:RingImages} must produce a cascade of damped oscillations on long baselines  (see, e.g., Fig.~5 of \cite{JohnsonLupsasca2020}).
Given that the $(n+1)^\text{th}$ photon ring has width
\begin{align}
	\label{eq:WidthScaling}
	w_{n+1}(\varphi)\approx e^{-\gamma(\varphi)}w_n(\varphi)
	\approx e^{-n\gamma(\varphi)}w_1(\varphi),
\end{align}
the $n^\text{th}$ subring ought to dominate the signal in the range
\begin{align}
	\label{eq:RingRange}
	\frac{1}{w_{n-1}}\ll u\ll\frac{1}{w_n},
\end{align}
in which the $(n-1)^\text{th}$ ring has already been resolved out, but the $(n+1)^\text{th}$ ring (whose flux is $\sim e^{-\gamma}$ times weaker) has yet to take over.
Hence, we expect that for large $n$, the visibility amplitude in the range \eqref{eq:RingRange} must adopt a universal form \eqref{eq:UniversalAmplitude} fixed by the $n^\text{th}$ ring diameter $d_\varphi^{(n)}$.

This expectation has been confirmed by simple models \cite{GLM2020,Paugnat2022,Vincent2022,CardenasAvendano2022} for which these statements already hold to very good approximation starting with the $n=2$ ring, as expected since its typical width $w_2\lesssim0.1M$ and diameter $d\sim10M$ correspond to a width-to-diameter ratio $\lesssim1\%$ suitable for an expansion in $w_2/d\ll1$.
Indeed, \cite{GLM2020} found that the diameter $d_\varphi^{(2)}$ of the $n=2$ ring image could be inferred from its visibility amplitude in the range \eqref{eq:RingRange}.

In general, the $n^\text{th}$ ring must lie within the $n^\text{th}$ ``lensing band'': an exponentially narrow (in $n$) region of the image plane that is  fully determined by the Kerr metric \cite{GLM2020,CardenasAvendano2022}.
Within this band, however, it generically has some width.
Two important comments are now in order:
\begin{enumerate}
	\item We reiterate that any ring of finite width does not have a unique, well-defined image diameter $d_\varphi$, but rather a range of diameters that extends from a minimum diameter (between its inner boundaries) to a maximum diameter (between the outer ones).
	That is, the image diameter is only defined up to a precision of order the ring width $w$ (Fig.~\ref{fig:RingDiameters}).
	Yet, the corresponding visibility in the universal regime \eqref{eq:UniversalRegime} does seem to pick out a unique periodicity---so how can this be?
	A resolution to this puzzle is partly that the exact periodicity of the ringing in the universal visibility \eqref{eq:UniversalAmplitude} varies with the baseline length $u$ within the regime \eqref{eq:UniversalRegime}.
	That is, the precise value of the diameter $d_\varphi^{(n)}(u)$ depends on the choice of baseline window from which it is inferred \cite{Paugnat2022}.
	\item[] For a thin $n=2$ ring, the image diameters vary within a narrow range of $\lesssim1\%$ (Fig.~\ref{fig:RingDiameters}), but such variation could be detected at the microarcsecond scale accessible on 230\,GHz Earth-Moon baselines.
	When observing near $u\sim300\,$G$\lambda$, for instance, a unique periodicity emerges and yields a sharp $n=2$ ring diameter $d_\varphi^{(2)}(u)$, but this answer varies with the baseline length $u$ within the regime \eqref{eq:RingRange}.
	As longer baselines are sampled, higher-frequency components of the ring are progressively picked up and larger image gradients within its intensity profile are increasingly resolved.
	Intuitively, then, the inferred diameter $d_\varphi^{(2)}$ receives  contributions from image diameters connecting points across the ring's profile where the derivative of the intensity is greater.
	As a result, the inferred diameter $d_\varphi^{(2)}(u)$ may exhibit a slight but still noticeable drift in $u$.
	\item We also emphasize that the universal signature \eqref{eq:UniversalVisibility}--\eqref{eq:UniversalAmplitude} is only present when the universal regime \eqref{eq:UniversalRegime} exists.
	That is, the $n^\text{th}$ photon ring produces its characteristic periodic ringing only within the range \eqref{eq:RingRange}.
	All the photon rings have a diameter $d\sim10M$, which for 230\,GHz observations of M87* corresponds to an angular size of $d\sim40\,\mu$as, and hence to a radial periodicity $\Delta u\sim1/d\approx5\,$G$\lambda$ \cite{EHT2019a}.
	As such, the number $N_n\approx(\Delta u_n)d$ of periods (or ``hops'') of the visibility amplitude within the regime \eqref{eq:RingRange} of width $\Delta u_n=1/w_n-1/w_{n-1}$ is $N_2\sim100$, a sufficient number to obtain a good estimate of the periodicity $\Delta u=1/d_\varphi^{(2)}$.
	The scaling \eqref{eq:WidthScaling} of the ring widths implies a scaling $\Delta u_{n+1}\approx e^\gamma\Delta u_n$ of the baseline windows \eqref{eq:RingRange}, so each ring produces an exponentially growing number $N_{n+1}\approx e^\gamma N_n$ of hops in the range in which it dominates the signal.
	Therefore, every $n\ge2$ ring produces sufficiently many hops to enable an estimate of its diameter.
\end{enumerate}

\subsection{Predicted interferometric shape of the higher (\texorpdfstring{$n\ge2$}{n>1}) photon rings}

The first property of Kerr lensing described in Sec.~\ref{sec:RingImages} (namely, the exponential demagnification of successive subrings) guarantees a small width-to-diameter ratio for all the $n\ge2$ subrings.
As a result, their diameter $d_\varphi^{(n)}$ is well-defined in the image, up to minute variations of order $w_n/d\ll1$ (with $w_2/d\sim1\%$ and the higher ratios $w_n/d$ exponentially suppressed by factors of $\sim e^{-\gamma}$).

Moreover, this guarantees---for each $n\ge2$ subring---the existence of a wide range of baselines in the ``universal regime'' \eqref{eq:UniversalRegime}.
In the regime \eqref{eq:WidthScaling} dominated by the $n^\text{th}$ ring, the visibility amplitude takes a ``universal form'' \eqref{eq:UniversalAmplitude} that is fixed by the ring diameter $d_\varphi^{(n)}$, and which extends over sufficiently many hops for its periodicity---and hence $d_\varphi^{(n)}$---to be precisely inferred.
This interferometrically measured diameter $d_\varphi^{(n)}(u)$ may vary slightly with the precise choice of baseline $u$ within the range \eqref{eq:UniversalRegime}, but this variation is again limited to variations of order $w_n/d\ll1$.

The outstanding question that remains is then: what is the interferometric shape of the $n\ge2$ rings?
Or, more precisely: does general relativity make a prediction for the projected diameter $d_\varphi^{(n)}$?

The second property of Kerr lensing described in Sec.~\ref{sec:RingImages} (namely, the exponential convergence of successive rings to the critical curve) answers in the affirmative: by \eqref{eq:Convergence}, the rings $\mathcal{C}_n$ converge to the critical-curve shape $\tilde{\mathcal{C}}=\tilde{C}_\infty$.
In other words, as the large-$n$ rings become increasingly narrow curves, their projected position functions tend to that of the critical curve, $\tilde{f}(\varphi)=\frac{1}{2}\tilde{d}_\varphi+\tilde{C}_\varphi$.
In particular,
\begin{align}
	\lim_{n\to\infty}d_\varphi^{(n)}=\tilde{d}_\varphi,\quad
	\lim_{n\to\infty}C_\varphi^{(n)}=\tilde{C}_\varphi.
\end{align}
The analytic expression for the critical curve's projected position $\tilde{f}(\varphi)$ is investigated in \cite{GrallaLupsasca2020c}.
The exact formula is rather unwieldy, but it closely tracks the ``phoval'' shape
\begin{align}
	\tilde{f}(\varphi)\approx R_0&+\sqrt{R_1^2\sin^2{\varphi}+R_2^2\cos^2{\varphi}}\\
	&+(X-\chi)\cos{\varphi}+\arcsin(\chi\cos{\varphi}),\notag
\end{align}
to better than 1 part in $10^5$ for most black hole spins and observer inclinations, with the largest deviation from this functional form reaching 1 part in $10^3$ in the extremal limit $a\to M$ for an equatorial observer, when the critical curve is least circular and develops a vertical edge \cite{Gralla2018}.

The five parameters $R_0$, $R_1$, $R_2$, $X$ and $\chi$ in the phoval family of shapes admit a simple geometric interpretation.
The offset $X$ accounts for a spin-dependent,  translation of the centroid of the critical curve relative to Bardeen's Cartesian coordinate system $(\alpha,\beta)$.
Together with the parameter $\chi\in[-1,1]$, which is necessary to reproduce the asymmetry of the high-spin, high-inclination critical curve, it only enters into the projected centroid of $\tilde{\mathcal{C}}$,
\begin{align}
	\tilde{C}_\varphi\approx(X-\chi)\cos{\varphi}+\arcsin(\chi\cos{\varphi}).
\end{align}
The projected diameter thus takes the functional form
\begin{align}
	\label{eq:CriticalCurveDiameter}
	\frac{\tilde{d}_\varphi}{2}\approx R_0+\sqrt{R_1^2\sin^2{\varphi}+R_2^2\cos^2{\varphi}},
\end{align}
controlled by three characteristic radii $R_0$, $R_1$ and $R_2$.
When $R_1=R_2=0$, this describes a curve of constant radius $R_0$, such as a perfect circle, which is the shape of $\tilde{\mathcal{C}}$ for an on-axis observer at any spin or for any observer at zero spin.
When $R_0=0$ instead, \eqref{eq:CriticalCurveDiameter} describes a perfect ellipse with axes of length $R_1$ and $R_2$, which is exactly the shape of $\tilde{\mathcal{C}}$ for all spins and low observer inclinations, or equivalently for all inclinations at small spin \cite{GrallaLupsasca2020c}.

Based on the reasoning laid out above, \cite{GLM2020} conjectured that the projected diameter of the $n\ge2$ rings of M87* (which due to its jet orientation is believed to be observed at a relatively low inclination of $\theta_{\rm o}\approx17^\circ$ \cite{EHT2019a}) ought to follow the four-parameter functional form of a ``circlipse''
\begin{align}
	\label{eq:Circlipse}
	\frac{d_\varphi^{(n)}}{2}\approx R_0+\sqrt{R_1^2\sin^2(\varphi-\bar{\varphi})+R_2^2\cos^2(\varphi-\bar{\varphi})},
\end{align}
where the additional offset angle $\bar{\varphi}$ is meant to account for the uncertain image-plane orientation of the projected black hole spin (and in practice, the low-$n$ subrings may also appear rotated relative to the critical curve).

As checked in \cite{GLM2020,Paugnat2022,Vincent2022} for multiple astrophysical source profiles around a Kerr black hole, the visibility amplitude in the regime \eqref{eq:UniversalRegime} dominated by the $n=2$ ring really does follow the universal form \eqref{eq:UniversalAmplitude}, with a projected $n=2$ ring diameter $d_\varphi^{(2)}$ following the circlipse shape \eqref{eq:Circlipse}.
We may therefore regard \eqref{eq:Circlipse} as a generic prediction for the interferometric signature of the $n\ge2$ rings that follows from the Kerr hypothesis.
As argued in \cite{GLM2020,Paugnat2022,Vincent2022,CardenasAvendano2022}, a measurement of the interferometric $n=2$ ring diameter $d_\varphi^{(2)}(u)$ on long Earth-space baselines could deliver a stringent test of strong-field general relativity.

\subsection{Predicted interferometric shape of the first (\texorpdfstring{$n=1$}{n=1}) photon ring}

Having reviewed the prediction for the interferometric shape of the $n\ge2$ photon rings, we now turn to the first photon ring, for which an analogous prediction has so far been lacking.
In large part, this is because the preceding discussion breaks down for the first $n=1$ subring:
\begin{enumerate}
    \item Because of its significant width $w_1\sim M$ and large width-to-diameter ratio $w_1/d\sim10\%$, the first ring really lacks a sharply defined diameter $d_\varphi^{(1)}$ in image space: at a given angle $\varphi$ around the image, it has a wide range of possible diameters (Fig.~\ref{fig:RingDiameters}).
    Moreover, unlike the higher-$n$ rings, the $n=1$ ring is not yet strongly constrained to track the critical curve shape $\tilde{\mathcal{C}}$.
    As a result, there is no theoretical prediction for its image diameter---more than that, it is not even clear how such a diameter could be precisely defined from the $n=1$ ring image.
    \item Relatedly, even if we were able to define the $n=1$ ring diameter $d_\varphi^{(1)}$ in image space and then derive a prediction for it, we would have no reason to expect its visibility to adopt the universal form \eqref{eq:UniversalVisibility}: first, because this form was derived in a leading-order expansion in $w/d\ll1$ and may receive significant corrections when $w_1/d\sim10\%$, and second, because the $n=1$ ring typically fails to exhibit a ``universal regime'' in which it dominates the signal, since the range \eqref{eq:UniversalRegime} closes off for thick rings with $w/d\gtrsim10\%$ (see App.~\ref{app:NoUniversalRegime} and App.~C of \cite{Paugnat2022} for more details).
\end{enumerate}

To get around all these issues, we propose to \textit{define} an interferometric diameter $d_\varphi^{(1)}(u)$ from the periodicity of the visibility amplitude in the baseline range where the $n=1$ ring dominates, namely
\begin{align}
	\label{eq:FirstRingRange}
    \frac{1}{d}<u\lesssim\frac{1}{w_1}.
\end{align}
We reiterate that this range is typically too narrow to contain a universal regime \eqref{eq:UniversalRegime}.
For observations of M87* at 230\,GHz, this range usually stretches from $\gtrsim25\,$G$\lambda$ to $\lesssim100\,$G$\lambda$, and therefore contains only $\sim15$ ``hops'' of periodicity $\Delta u\sim5\,$G$\lambda$.
Nevertheless, even a handful of hops is already enough to estimate a periodicity, and thence infer a diameter $d_\varphi^{(1)}$.

Two final questions remain.
First, we have no reason to expect the visibility amplitude in the (non-universal) $n=1$ regime \eqref{eq:FirstRingRange} to take the universal form \eqref{eq:UniversalAmplitude}, so there is no obvious functional form to fit to the visibility.
How then can we best extract its periodicity?

Second, assuming an interferometric diameter $d_\varphi^{(1)}$ can be extracted from the visibility amplitude, what form should its angle-dependence take?
Since this diameter would have no clear connection to any precise feature in the image, it is perhaps not evident what to expect.

To tackle the first problem, we note that the universal visibility amplitude \eqref{eq:UniversalAmplitude} can be generalized to \cite{Paugnat2022}
\begin{align}
	\label{eq:AmplitudeFit}
	\ab{V(\mathbf{u})}&\approx\sqrt{\pa{A_\varphi^{\rm L}}^2+\pa{A_\varphi^{\rm R}}^2+2A_\varphi^{\rm L}A_\varphi^{\rm R}\sin\pa{2\pi d_\varphi u}},
\end{align}
where, instead of decaying like $\sqrt{u}$, the angle-dependent functions $A_\varphi^{\rm L/R}$ may become general functions of $u$,
\begin{align}
	\label{eq:Envelopes}
	A_\varphi^{\rm L/R}(u)=\frac{e_{\rm upper}(u)\pm e_{\rm lower}(u)}{2}.
\end{align}
Here, $e_{\rm upper}(u)$ and $e_{\rm lower}(u)$ respectively correspond to the upper and lower envelopes of the function \eqref{eq:AmplitudeFit}, which oscillates between these envelopes with periodicity $d_\varphi$.
As shown in \cite{Paugnat2022,CardenasAvendano2022}, fitting a ringing signal to \eqref{eq:AmplitudeFit} is a more robust method for inferring its periodicity, even when it takes the universal form \eqref{eq:UniversalAmplitude}.
Mathematically speaking, we know of no reason why it should always be possible to fit a generic ringing visibility to the functional form \eqref{eq:AmplitudeFit}, but in practice we find that it is sufficiently general to work (see Sec.~3.2.2 of \cite{Paugnat2022} for more discussion).

As for the second question, the simplest guess is that the interferometrically defined $n=1$ ring diameter $d_\varphi^{(1)}$ still follows the circlipse shape \eqref{eq:Circlipse}, at least for the low observer inclinations relevant to M87* observations.
At this stage of the discussion, this is merely a conjecture which may not necessarily be correct.
As we will show in the remainder of the paper, however, it does turn out to be true (to about 1 part in $10^3$) in a wide range of simple phenomenological models of M87*.

As such, we may also regard \eqref{eq:Circlipse} as a prediction from the Kerr hypothesis for the interferometric shape of the $n=1$ ring, and its measurement could provide a precise probe of general relativity in the strong-field regime.

\subsection{Comparison with the shadow and critical curve}

In certain highly fine-tuned scenarios, the photon ring and its subring substructure are not present in black hole images.
This happens, for instance, when the black hole is immersed in a spherically symmetric accretion inflow: in that case, the observational appearance of the source consists of a bright ring that encircles a central brightness depression whose boundary precisely coincides with the critical curve---an effect known as the ``black hole shadow'' \cite{Falcke2000,Narayan2019}.
In such a scenario, measuring the shadow---the shape of the central brightness deficit---yields a direct measurement of the critical curve $\tilde{\mathcal{C}}$, and hence of the black hole geometry.
Indeed, the shape of $\tilde{\mathcal{C}}$ depends only on the black hole mass $M$, its spin $a$, and the inclination $\theta_{\rm o}$ of the observer, and these three parameters can be directly recovered from the three radii $(R_0,R_1,R_2)$ that parametrize the projected diameter \eqref{eq:CriticalCurveDiameter} of $\tilde{\mathcal{C}}$.

Unfortunately, such an astrophysical scenario does not seem to be relevant for either M87* or Sgr A* \cite{Gralla2019,JohnsonLupsasca2020,GrallaLupsasca2020a,Chael2021}, which are instead expected to present the photon ring structure described in Sec.~\ref{sec:RingImages}.
Thus, the analytically known critical curve, which directly encodes the black hole parameters $(M,a,\theta_{\rm o})$, is likely not observable in itself.
On the other hand, the photon rings, which are in principle observable, do not have an analytically predicted shape that encodes $(M,a,\theta_{\rm o})$.
Rather, their appearance is not entirely fixed by the Kerr geometry, but instead varies with the astrophysical details of the emitting source: indeed, two black holes with the same mass and spin, observed from the same inclination, can nonetheless display photon rings of noticeably different shapes if their emission differs \cite{GLM2020}.

In other words, while the shape of the photon rings does track that of the critical curve, in the sense that their projected diameters follow the same functional form \eqref{eq:CriticalCurveDiameter}--\eqref{eq:Circlipse}, the radii parametrizing these functions differ in their interpretation: in \eqref{eq:CriticalCurveDiameter}, they can be mapped back to $(M,a,\theta_{\rm o})$, whereas in \eqref{eq:Circlipse}, this map itself depends on the source, with the astrophysical dependence vanishing as $n\to\infty$.
Hence, measuring the projected diameter \eqref{eq:Circlipse} of the $n^\text{th}$ photon ring gives stronger constraints on $(M,a,\theta_{\rm o})$ the higher $n$ is.
For $n=2$, these can likely be inferred within a few percent, but less precisely for $n=1$.

\subsection{Summary of the predicted first ring shape}
\label{subsec:Summary}

To summarize, we predict that in the baseline range \eqref{eq:FirstRingRange} dominated by the $n=1$ ring, the visibility amplitude of the (time-averaged) image of a Kerr black hole displays a characteristic ringing with angle-dependent periodicity $\Delta u=1/d_\varphi^{(1)}$, where $d_\varphi^{(1)}$ follows the functional form \eqref{eq:Circlipse} to high precision.
This naturally extends a previous \cite{GLM2020} and similar prediction for the higher-$n$ rings to the first and most easily accessible $n=1$ subring.

In contrast to the $n\geq2$ rings, for which $d_\varphi^{(n)}$ may be associated with the diameter of the $n^\text{th}$ ring in the image, the $n=1$ ring diameter $d_\varphi^{(1)}$ is defined purely interferometrically and lacks a sharp image-space interpretation.

The diameter $d_\varphi^{(1)}$ may be extracted from the visibility amplitude by fitting the latter to \eqref{eq:AmplitudeFit} and finding the best-fitting envelopes \eqref{eq:Envelopes} and circlipse shape \eqref{eq:Circlipse}. 
The three circlipse radii $(R_0,R_1,R_2)$ loosely track the black hole parameters $(M,a,\theta_{\rm o})$, but their precise relation is not robust and depends on the astrophysics of the source.

\section{Phenomenological source model}
\label{sec:Model}

\begin{figure}
    \centering
    \includegraphics[width=\columnwidth]{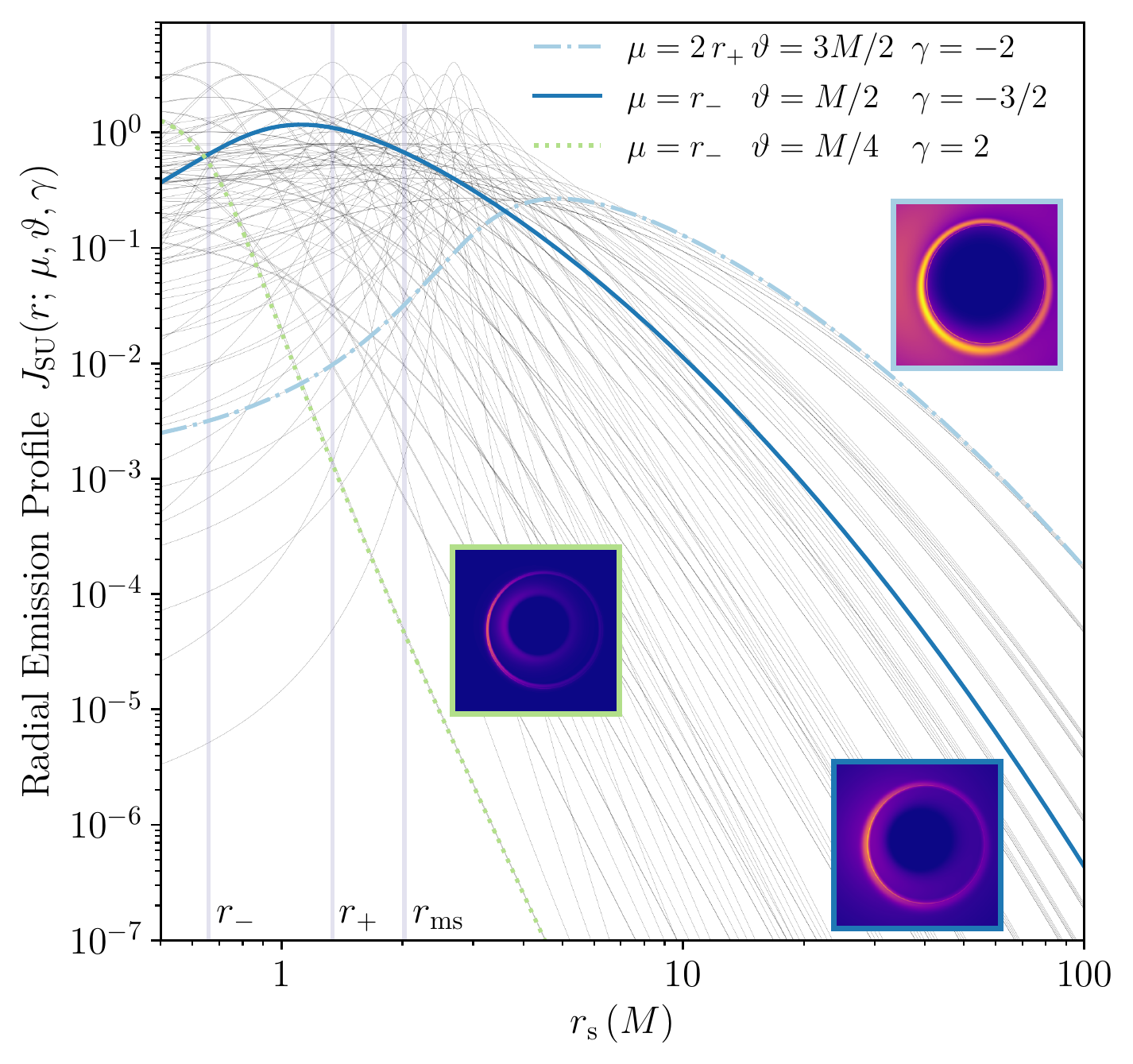}
    \caption{Radial emission profiles \eqref{eq:RadialProfile} considered in this work (with parameters listed in Table~\ref{tbl:Parameters}).
    These profiles range from emission that peaks inside the horizon and then decays rapidly outside (dotted green), to emission that peaks past the innermost stable circular orbit (with ISCO radius $r_{\rm ms}$) and decays very slowly thereafter (dash-dot blue).
    The radii of the outer and inner horizons are denoted by $r_\pm=M\pm\sqrt{M^2-a^2}$ and indicated with vertical lines.
    The profile depicted with a solid blue line was considered in \cite{GLM2020}.
    It is broadly consistent with the 2017 EHT observations of M87* on Earth-size baselines and qualitatively similar to time-averaged images of state-of-the-art general-relativistic magnetohydrodynamic (GRMHD) simulations (see, e.g., Fig.~1 of \cite{JohnsonLupsasca2020}).
    The insets display the images produced by each of these three highlighted profiles.}
    \label{fig:Profiles}
\end{figure}

\begin{table}
	\centering
	\begin{tabular}{c c}
 	\hline
    \hline
	Johnson SU Parameter & Values \\
	\hline
	$\mu$ & $r_-,r_+/2,r_+,3r_+/2,2r_+$ \\
	$\gamma$ & $-2,-1,0,1,2$ \\
	$\vartheta/M$ & $0.25, 0.5, 1.0, 1.5$ \\
 	\hline
    \hline
	\end{tabular}
	\caption{Values of the parameters considered in our survey over the 100 radial emission profiles \eqref{eq:RadialProfile} shown in Fig.~\ref{fig:Profiles}.
	The outer/inner event horizon radii are $r_\pm=M\pm\sqrt{M^2-a^2}$.}
	\label{tbl:Parameters}
\end{table}

Having formulated a prediction for the interferometric shape of the $n=1$ ring, we now wish to test whether it does indeed hold in a simple phenomenological model of M87*.
In this section, we give a lightning review of the source model introduced in \cite{GLM2020,Chael2021,Paugnat2022,CardenasAvendano2022}.
Then, we use our Adaptive Analytical Ray Tracing code \texttt{AART} \cite{CardenasAvendano2022}---which exploits the integrability of light propagation in the Kerr spacetime---to compute high-resolution black hole images of this model, together with their corresponding visibilities accessible on long space-ground baselines.

We model the source as a stationary, axisymmetric, equatorial disk composed of emitters describing circular Keplerian orbits down to the radius $r_{\rm ms}$ of the innermost stable circular orbit (ISCO), past which they plunge into the hole following a prescription of Cunningham~\cite{Cunningham1975}.

To determine the observational appearance of the source at an image-plane position $(\alpha,\beta)$, we analytically trace the corresponding light ray back from the observer's image plane and into the emitting region, increasing the observed intensity $I_{\rm o}(\alpha,\beta)$ each time the ray intersects the accretion disk by an amount dictated by the local emissivity.
The full procedure is efficiently implemented in our relativistic ray tracing code \texttt{AART} \cite{CardenasAvendano2022}.

For completeness, we sketch the main ingredients of the calculation, referring the reader to \cite{CardenasAvendano2022} for the details of our implementation.
Effectively, we compute
\begin{align}
	I_{\rm o}(\alpha,\beta)=\sum_{n=0}^{N(\alpha,\beta)-1}\zeta_n\cdot g^3\pa{r_{\rm s}^{(n)},\alpha}I_{\rm s}\left(r^{(n)}_{\rm s}\right),
\end{align}
where $r_{\rm s}^{(n)}=r_{\rm s}^{(n)}(\alpha,\beta)$ denotes the (analytically known) equatorial radius at which a ray intersects the equatorial plane for the $(n+1)^\text{th}$ time on its backward trajectory from image-plane position $(\alpha,\beta)$, up to a total number $N(\alpha,\beta)$ along its maximal extension.
Meanwhile, $g$ is a redshift factor (which is determined by the motion of the emitters and also known analytically), $I_{\rm s}(r)$ is a radial emission profile, and $\zeta_n$ is a ``fudge'' factor, which we assume to be equal to 1 for $n=0$, and $1.5$ for $n\geq1$.

The inclusion of this factor is meant to account for the (otherwise neglected) effects of the disk's geometrical thickness.
It improves the qualitative agreement between images of this equatorial model and the time-averaged images obtained from state-of-the-art general-relativistic magnetohydrodynamic (GRMHD) simulations~\cite{Chael2021,Vincent2022}.

We consider a family of radial emission profiles derived from Johnson's Standard Unbounded (SU) distribution,
\begin{align}
	\label{eq:RadialProfile}
	I_{\rm s}(r_{\rm s})=J_{\rm{SU}}(r_{\rm s};\mu,\vartheta,\gamma)
	\equiv\frac{e^{-\frac{1}{2}\br{\gamma+\arcsinh\pa{\frac{r_{\rm s}-\mu}{\vartheta}}}^2}}{\sqrt{\pa{r_{\rm s}-\mu}^2+\vartheta^2}},
\end{align}
where the  parameters $\mu$, $\vartheta$, and $\gamma$ respectively control the location of the profile's peak, its width, and the profile asymmetry \cite{Paugnat2022}.
In our survey over emission profiles, we examine the same set of parameters as in \cite{Paugnat2022}, which we list in Table~\ref{tbl:Parameters}.
We display the corresponding profiles in Fig.~\ref{fig:Profiles}, which shows that these values span a wide range of possible emissivities.
In particular, our survey includes a profile (solid blue line in Fig.~\ref{fig:Profiles}) whose corresponding image is directly comparable to the time-averaged image in Fig.~1 of \cite{JohnsonLupsasca2020}, for which the parameters in the underlying GRMHD simulation were chosen to ensure consistency with the 2017 EHT observations of M87*. 

\section{Survey over emission profiles}
\label{sec:Results}

Using \texttt{AART} \cite{CardenasAvendano2022}, we perform a parameter survey over the emission profiles in Table~\ref{tbl:Parameters} and Fig.~\ref{fig:Profiles}, enabling us to verify how well our prediction for the interferometric shape of the $n=1$ ring holds in our model.

For each of the 100 considered emission profiles, we ray trace high-resolution images and compute the associated visibility amplitudes on very long baselines with \texttt{AART}---we assume throughout a black hole spin of $a/M=94\%$ and an observer inclination $\theta_{\rm o}=17^\circ$, but we expect our conclusions to hold more generally at low inclinations.

Then, following the procedure introduced in \cite{Paugnat2022}, we determine the functional forms \eqref{eq:AmplitudeFit}--\eqref{eq:Envelopes} that best fit the visibility amplitude in a given baseline range, allowing us to extract a characteristic periodicity for its ringing and thereby infer a projected diameter $d_\varphi$.

As discussed in \cite{GLM2020,Paugnat2022}, on baselines of length $\sim u$, the functional form \eqref{eq:AmplitudeFit} is approximately invariant under shifts $d_\varphi\to d_\varphi+k/u$ for integer $k$, creating a discrete degeneracy in the inferred diameter $d_\varphi$.
In principle, this degeneracy may be broken by counting the exact number of hops from $u$ back to the origin $u=0$, which would fix the radial periodicity $\Delta u$ of the ringing, and hence $d_\varphi$.
In practice, this is not possible if we can only sample a fixed baseline window far from the origin.
Instead, we fit $d_\varphi$ at multiple baseline angles $\varphi$ simultaneously, so as to obtain the best global fit $d_\varphi^{\rm obs}$ for the interferometric diameter.
This multi-fit procedure is explained in Sec.~3.2 of \cite{Paugnat2022}.

We carry out the fit on four different baseline windows of width $30\,$G$\lambda$, and assess how well the resulting interferometric diameters $d_\varphi^{\rm obs}(u)$ match the prediction \eqref{eq:Circlipse} in each of the windows as a function of the emission profile.
In realistic observations, the sampled baseline window would likely vary in size and location, so this simulates a somewhat idealized experiment.
We chose a fixed width of $30\,$G$\lambda$ to ensure that the baseline windows contain $\sim5$ hops in the visibility amplitude.

The first three baseline windows that we examine are $[40,70]\,$G$\lambda$, $[50,80]\,$G$\lambda$, and $[70,100]\,$G$\lambda$.
These often lie in the range \eqref{eq:FirstRingRange} dominated by the $n=1$ ring, though sometimes they may also fall into a region where the $n=0$ image still contributes significant power.

Meanwhile, the fourth window spans the much longer baselines $[285,315]\,$G$\lambda$, which were investigated in \cite{GLM2020} and typically fall into a universal regime \eqref{eq:UniversalRegime} dominated by the $n=2$ ring.
For some profiles, however, these baselines fall into a transition regime between the range \eqref{eq:FirstRingRange} dominated by the $n=1$ ring and the universal $n=2$ range \eqref{eq:UniversalRegime}.
As explained in Sec.~4 of \cite{Paugnat2022}, in such cases the inferred diameter $d_\varphi^{\rm obs}$ may belong to neither ring, as the visibility amplitude receives significant contributions from both of them, resulting in interference.
Or, it may happen that $d_\varphi^{\rm obs}$ measures the $n=1$ ring diameter $d_\varphi^{(1)}$ at some angles $\varphi$, and the $n=2$ diameter $d_\varphi^{(2)}$ for others.

In principle, all these baselines are within reach of an eccentric orbiter whose Earth perigee is at $\sim4.5\times10^4\,$km and its apogee at $\sim4\times10^5\,$km, provided it is equipped with receivers capable of observing on multiple frequency bands between $83\,$GHz and $345\,$GHz.
Indeed, a baseline extending from Earth to such an orbiter could sample the visibility from $\sim12\,$G$\lambda$ (shortest baseline with the lowest frequency) to $\sim455\,$G$\lambda$ (largest baseline with the highest frequency).

\begin{figure}
    \centering
    \includegraphics[width=\columnwidth]{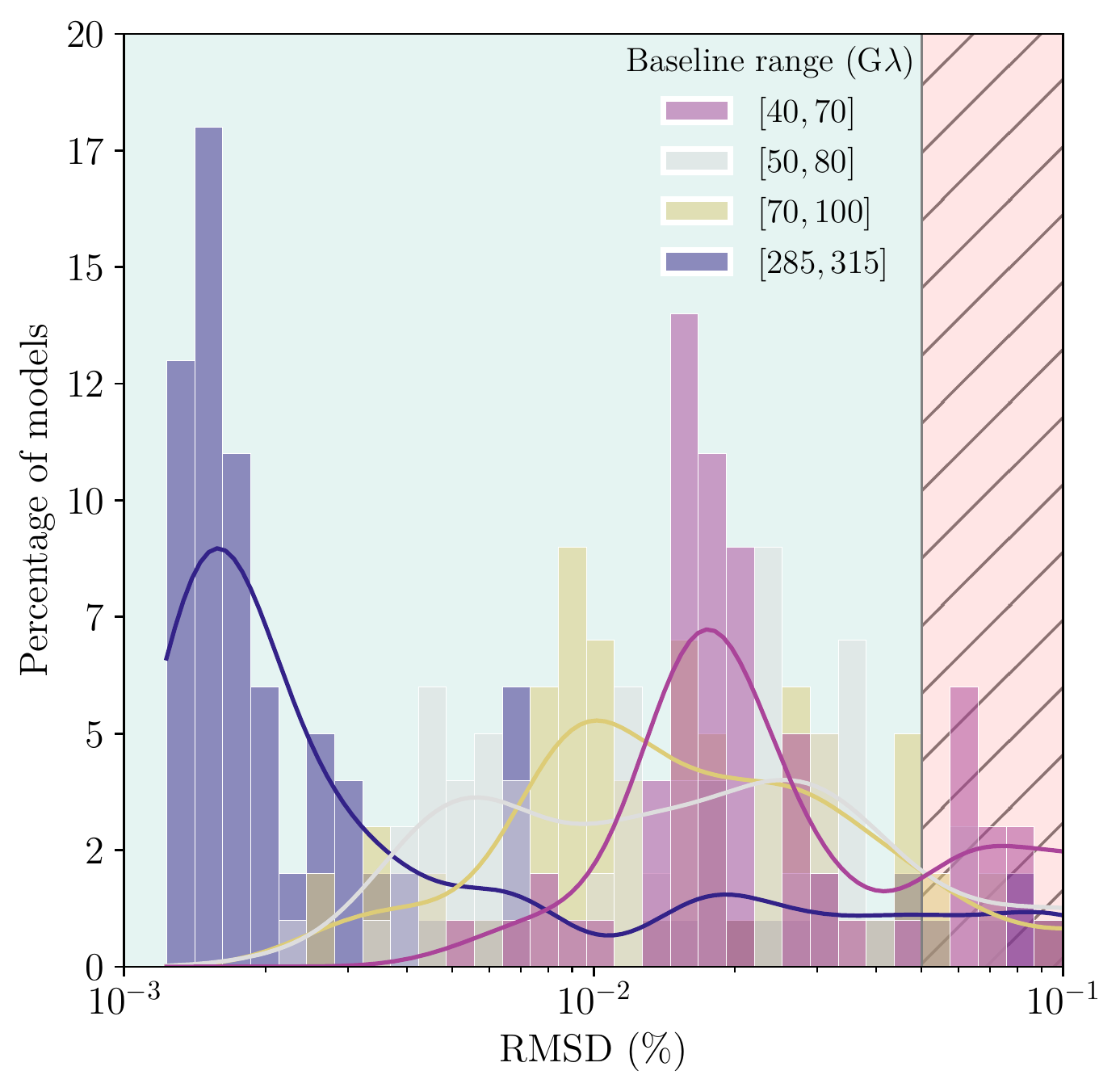}
    \caption{The normalized root-mean-square deviation (RMSD) distributions of the fits for four different windows of size $30\,$G$\lambda$ with overlaying kernel density estimations.
    The percentage of models for which we get an acceptable fit (RMSD $\ge0.05\%$) are $53\%$, $70\%$, $72\%$, and $85\%$ for the windows $[40,70]\,$G$\lambda$, $[50,80]\,$G$\lambda$, $[70,100]\,$G$\lambda$, and $[285,315]\,$G$\lambda$, respectively.
    For the baseline window $[70,100]\,$G$\lambda$, we display the best fit in the top panels of Fig.~\ref{fig:BestWorstFit}, while the worst fit (which still has an RMSD $\leq0.05\%$) is shown in the bottom panels of Fig.~\ref{fig:BestWorstFit}.}
    \label{fig:ModelHistograms}
\end{figure}

Given an observed diameter $d_\varphi^{\rm obs}$, we assess the quality of its best fit $d_\varphi^{\rm GR}$ to the circlipse shape \eqref{eq:Circlipse} by computing the ring-averaged normalized root-mean-square deviation
\begin{align}
     {\rm RMSD}=\frac{\sqrt{\av{\pa{d_\varphi^{\rm obs}-d_\varphi^{\rm GR}}^2}_\varphi}}{\av{d_\varphi^{\rm GR}}_\varphi}.
\end{align}

The RMSD distributions resulting from the fits on the four different windows are shown in Fig.~\ref{fig:ModelHistograms}.
As expected, the shortest baseline window is the one where fitting a circlipse shape to the inferred diameter is hardest: out of the 100 profiles studied, 53 provided a RMSD $\le0.05\%$, which is our (conservative) cutoff.
The number of models for which the RMSD $\le0.05\%$ jumps to 85 for the longest baseline window considered.

\begin{figure*}[ht!]
    \centering
    \includegraphics[width=\textwidth]{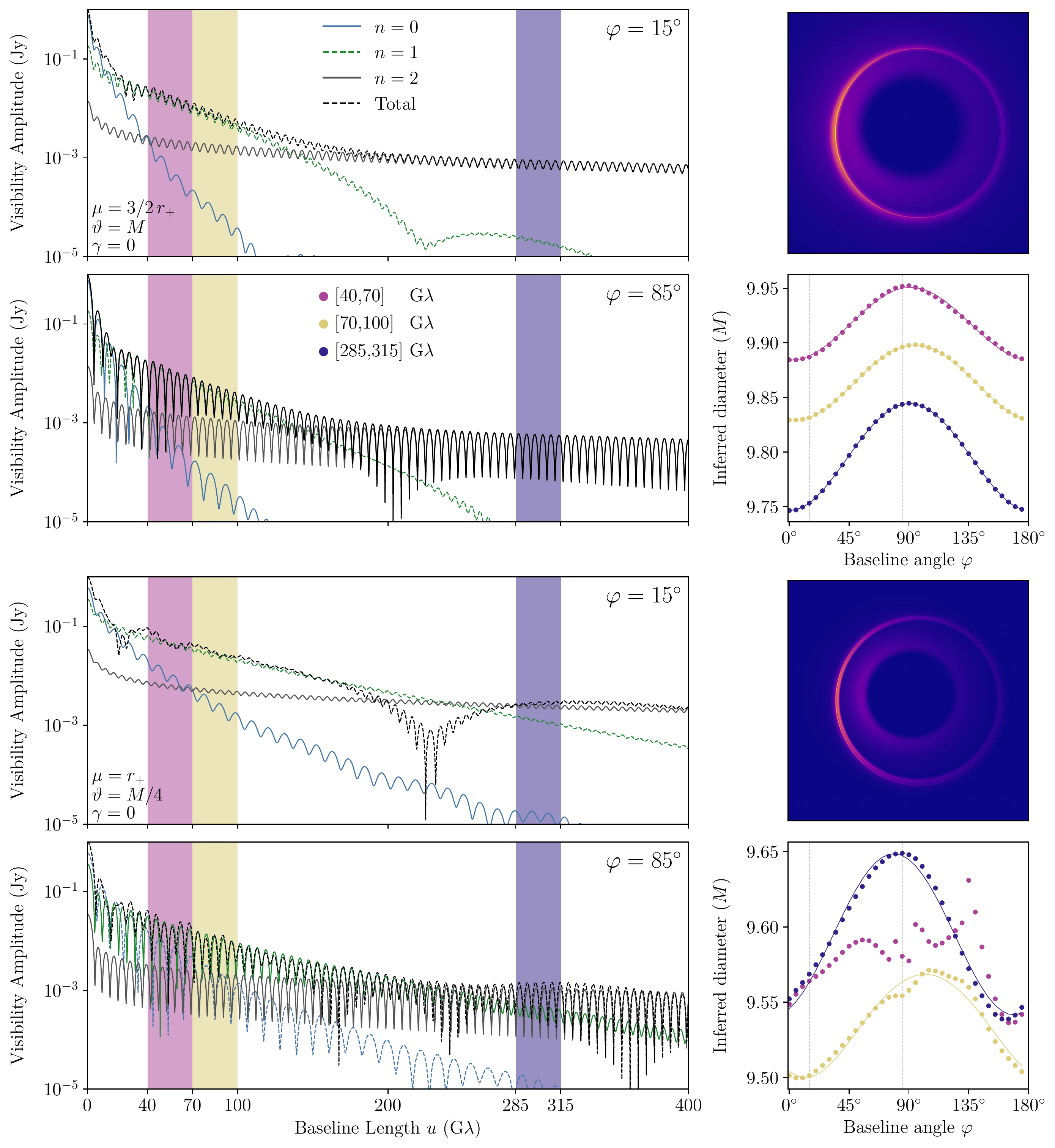}
    \caption{Visibility amplitudes at two different baseline angles ($\varphi=15^\circ$ and $\varphi=85^\circ$), together with corresponding images and inferred diameters $d_\varphi$, for two emission profiles from our survey (with Johnson SU parameters shown in the lower-left corner).
    The top two rows display the profile whose interferometric $n=1$ ring diameter $d_\varphi^{(1)}$ inferred from the periodicity of the visibility amplitude in the baseline window $[70,100]\,$G$\lambda$ (yellow) best fits our theoretical prediction \eqref{eq:Circlipse}, with an RMSD of $0.0025\%$.
    Conversely, the bottom two rows correspond to the profile in our survey with the worst fit to the circlipse shape \eqref{eq:Circlipse} that nonetheless has RMSD $\leq0.05\%$.
    For the best-fit profile, an excellent fit is also obtained in the other two baseline windows: RMSD of $0.012\%$ in $[40,70]\,$G$\lambda$ (magenta), and RMSD of $0.0016\%$ in $[285,315]\,$G$\lambda$ (purple).
    On the other hand, for the worst-fit profile, a good fit is not possible in the window $[40,70]\,$G$\lambda$ because it lies in the transition region between the regimes dominated by the $n=0$ and $n=1$ rings: at many baseline angles in this range, each of the two rings produces significant power, and their signals interfere in the visibility, whose periodicity then corresponds to neither's diameter.
    For each profile, we decompose the total visibility amplitude into the contributions it receives from each image layer labeled by $n$.
    This total amplitude (dashed black line) is the sum of all the subring contributions in image space, but not in Fourier space.}
    \label{fig:BestWorstFit}
\end{figure*}

Here, it is important to clarify that the inability to obtain good circlipse fits within a fixed baseline window for some of the emission profiles does not mean that the diameters of their rings fail to follow the circlipse shape \eqref{eq:Circlipse}.
In fact, for all of our profiles, each ring produces a clean ringing signature, as expected from our discussion in Sec.~\ref{sec:RingSignatures}.
This can be seen, for instance, in Fig.~\ref{fig:BestWorstFit}, where we display the total visibility amplitude---together with its decomposition into separate subring contributions---for two representative profiles: one with the best-fitting circlipse in the window $[70,100]\,$G$\lambda$ (top two rows), and another with the worst fit whose RMSD $\le0.05\%$ we nonetheless deemed acceptable (bottom two rows).
The difficulty, therefore, lies in whether the ring diameters can be extracted from the total visibility amplitude, which is of course the only one that is observable in experiment.

Consistent with \cite{GLM2020,Paugnat2022}, we find that for the $n=2$ ring, one may always obtain a good fit to the circlipse shape \eqref{eq:Circlipse} from the total visibility amplitude, though this may require going to extremely long baselines (sometimes as far as $1000\,$G$\lambda$ \cite{Paugnat2022}).
This is only possible to do while still remaining in the universal regime \eqref{eq:UniversalRegime} dominated by the $n=2$ ring because the width of this range is so large.

By contrast, while the $n=1$ ring image by itself also produces a clean interferometric ringing, this signature only dominates the total visibility in the comparably much narrower range \eqref{eq:FirstRingRange}.
As a result, in some models, there may not exist a single baseline range in which the $n=1$ ring dominates at every baseline angle $\varphi$ (so that its full angle-dependent diameter $d_\varphi^{(1)}$ may be extracted).

By fixing a baseline length (window size) and angle, we may thus obtain a visibility amplitude that can be either dominated by a single ring or can show comparable power in multiple rings, as illustrated in Fig.~\ref{fig:BestWorstFit}.
Here, the relevant factor is primarily the width of the profile, which determines the locations and widths of the ranges over which each ring dominates the signal by itself.

On longer baselines $\sim300\,$G$\lambda$, one may see interference between the $n=1$ and $n=2$ rings.
This does not happen for the top model in Fig.~\ref{fig:BestWorstFit}, for which the visibility amplitude is completely dominated by the $n=2$ ring in the purple window, resulting in an excellent circlipse fit.

For the bottom model, on the other hand, the emission profile---and hence the $n=1$ ring---are much narrower, so the $n=1$ range \eqref{eq:FirstRingRange} extends farther out and there is some interference between $n=1$ and $n=2$ in the purple window $[285,315]\,$G$\lambda$, resulting in a slightly worse circlipse fit.
Nonetheless, it is clear that going farther out to even longer baselines would further attenuate the power from the $n=1$ ring and lead to a better circlipse fit to the $n=2$ ring diameter in this model also.

Neither of these models displays a transition between measuring purely $d_\varphi^{(1)}$ or $d_\varphi^{(2)}$ at different $\varphi$, though this can also sometimes occur---particularly at higher inclinations---as shown in Fig.~6 of \cite{Paugnat2022}.

On shorter baselines $\lesssim100\,$G$\lambda$, a clean diameter can be more challenging to extract because of more frequent interference between the $n=0$ and $n=1$ rings.
When such interference occurs, the inferred diameter can differ more from that of a circlipse, which is expected since the $n=0$ ring is not constrained to closely follow that shape.

For the top model in Fig.~\ref{fig:BestWorstFit}, such interference never occurs in either the magenta window $[40,70]\,$G$\lambda$ nor the yellow window $[70,100]\,$G$\lambda$, which are always dominated by the $n=1$ ring and from which we can therefore extract diameters $d_\varphi^{(1)}$ with excellent circlipse fits.

For the bottom model in Fig.~\ref{fig:BestWorstFit}, in the yellow window $[70,100]\,$G$\lambda$, there are baseline angles (such as $\varphi=15^\circ$) where the $n=1$ ring dominates, but at other angles (such as $\varphi=85^\circ$) the $n=0$ ring retains significant power.
As a result, the circlipse fit is not as good at those angles.

For extremely narrow profiles, the first three image layers ($n=0$ through $n=2$) all consist of very thin rings, and one may even observe interference effects between all three---this is for instance the case for the bottom model in Fig.~\ref{fig:BestWorstFit} in the magenta window $[40,70]\,$G$\lambda$ at $\varphi=85^\circ$.
Even at other angles (such as $\varphi=15^\circ$) where the $n=0$ signal has relatively died out, the $n=1$ and $n=2$ rings still have comparable power, which explains why the circlipse fit in that window is so poor overall.

These angle-dependent effects ought to grow with the observer inclination, but we expect them not to pose an insuperable obstacle at the low-to-moderate inclinations $\lesssim30^\circ$ of likely relevance for M87*.
As for the black hole spin, higher spins increase the angular variation of $d_\varphi$ and should therefore facilitate a precise circlipse fit---we defer a more thorough investigation to future work.

\section{Resolving the photon ring}
\label{sec:Detection}

\begin{figure*}
    \centering
    \includegraphics[width=0.9\textwidth]{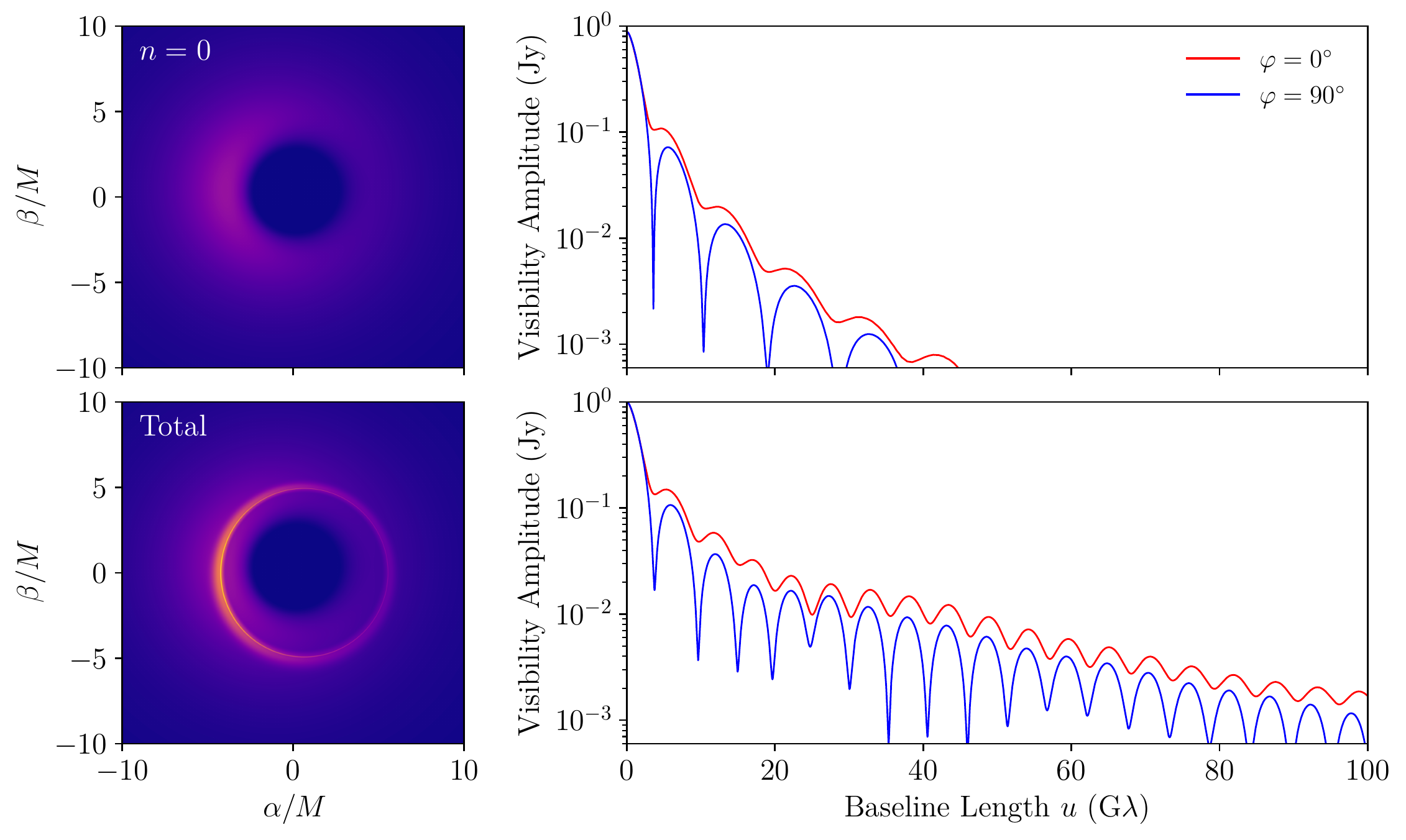}
    \caption{Gravity adds a distinctive ``stamp'' to black hole images and their interferometric visibilities.
    Time-averaged images of an M87* model with the strongly lensed light that produces the photon ring turned off (top left) and then back on (bottom left), together with their corresponding radio visibilities (right column).
    The interferometric signature of the photon ring is a characteristic ``ringing'' visibility amplitude at $\gtrsim20\,$G$\lambda$.
    The periodicity of this ringing at any given angle $\varphi$ in the baseline plane encodes the ring diameter $d_\varphi$ at the corresponding angle $\phi=\varphi$ in the image.}
    \label{fig:GravityStamp}
\end{figure*}

In this section, we discuss some of the implications of our results for future measurements of the $n=1$ ring, before concluding in Sec.~\ref{sec:Conclusion}.

There already exist multiple promising ways to detect the photon ring via its distinctive polarimetric signature \cite{Himwich2020} or characteristic pattern of autocorrelations \cite{Hadar2022,GrallaLupsasca2023}.
Here, we set aside these potential avenues for detection and focus exclusively on the complex visibility \eqref{eq:ComplexVisibility} dual to the image intensity, and, more precisely, its amplitude.

In that case, it seems likely that the first unambiguous observation of the photon ring will occur via a detection of its characteristic ``ringing'' in the visibility amplitude.
Moreover, this ringing will likely first be detected on the relatively shorter Earth-to-space baselines $u\lesssim100\,$G$\lambda$ most easily accessible to observations, where the first subring dominates the visibility.
Hence, our prediction \eqref{eq:Circlipse} for the angle-dependent diameter $d_\varphi^{(1)}$ of the $n=1$ ring, which may be inferred from the angle-dependent radial periodicity of the ringing in the visibility amplitude, is especially timely.

However, an important caveat is in order at this stage.
According to Fig.~\ref{fig:GravityStamp}, a ringing in the visibility amplitude on baselines of length $\sim20-40\,$G$\lambda$ does not by itself provide conclusive evidence for the presence of the photon ring.
After all, the $n=0$ image is also ring-like \cite{EHT2019a} (in line with theoretical expectations) and therefore it also produces a characteristic ringing on those baselines by itself (top row of Fig.~\ref{fig:GravityStamp}).
Naively, it seems necessary to probe longer baselines $\gtrsim40\,$G$\lambda$ to determine whether there really is a signature of the photon ring, that is, of strong gravity's stamp (bottom row of Fig.~\ref{fig:GravityStamp}).
For M87* observations at 230\,GHz, an interferometer with a space element appears to be indispensable to achieve the requisite baseline length.

Of course, a precise threshold past which the $n=0$ signal decays depends on the width of the $n=0$ ring in the image.
Likewise, the decay rate of the visibility amplitude in the regime \eqref{eq:FirstRingRange} dominated by the $n=1$ ring may well provide some information about its width.

It would be interesting to determine whether this width---and hence the Kerr-predicted demagnification factor $e^{-\gamma}$---may be recovered from the falloff rate of the visibility, that is, whether the envelope of the damped oscillations constrains the ring width.

We take some first steps in this direction in App.~\ref{app:ThickRings}, where we investigate the relation between the width $w$ of a Gaussian ring and the falloff rate $e^{-2\pi^2w^2u^2}$ of its visibility amplitude.
Generalizing such relations---if possible---would be interesting, since a measurement of $\gamma$ could yield a much-sought-after estimate of the black hole parameters, particularly its spin.

Finally, we note that a measurement of the predicted shape \eqref{eq:Circlipse} for the interferometric diameter of the $n=1$ ring would yield a consistency test of strong-field general relativity, since measuring a different diameter would be incompatible with Kerr hypothesis.
At the same time, a measurement of the expected circlipse shape would not necessarily discriminate between general relativity and alternative theories of gravity predicting the same shape.

On that note, \cite{Staelens2023} recently investigated the shape of the $n=2$ photon ring in modified theories of gravity.
They found that deviations from the circlipse shape were small unless the deviation from the Kerr geometry grew very large.
This conclusion merits re-evaluation in the context of the $n=1$ ring, whose deviations could perhaps be stronger.

\section{Conclusion}
\label{sec:Conclusion}

In this paper, we examined the shape of the first $n=1$ photon ring in time-averaged images of a simple model of M87* with an equatorial source around the black hole.

We found that, even though this first photon ring lacks a sharply defined diameter in the image domain (unlike the $n\ge2$ rings, which are much narrower and converge to the critical curve exponentially in $n$), it is nevertheless possible to define its angle-dependent projected diameter from the periodic ringing of its interferometric signature in visibility space, which is the observable that proposed extensions of the EHT to space will directly probe.

We showed in the context of our simple model of M87* that this interferometrically defined $n=1$ ring diameter follows the shape \eqref{eq:Circlipse} of a circlipse.
We therefore regard this as a prediction from strong-field general relativity (and its Kerr hypothesis) for the shape $d_\varphi^{(1)}$ of the first photon ring, which is most accessible to observation and will hopefully be measured soon.

We emphasize that the prediction of a circlipse shape \eqref{eq:Circlipse} for $d_\varphi^{(n)}$ is theoretically well-motivated when $n\ge2$.
The idea that the same shape would also describe the $n=1$ ring is a guess that we empirically observed to hold in our models.

The results of this work indicate that measuring $d_\varphi^{(1)}$ within a few percent of the critical curve diameter $\tilde{d}_\varphi$ is possible for several astrophysical profiles, and important factors affecting the accuracy of these measurements have been highlighted. 

As a final caveat: we have only studied time-averaged images and need to examine instantaneous snapshots with noise---both instrumental and astrophysical---to truly establish this as a robust prediction.
This work is now ongoing, and we expect to report results soon.
Overall, this line of research provides valuable insights into the interferometric structure of black hole images and lays the groundwork for future observations.
\vspace{10pt}

\acknowledgments

We thank Samuel Gralla and Daniel Marrone for their valuable comments.
We are grateful to Will and Kacie Snellings for their generous support, and A.C.-A. also acknowledges support from the Simons Foundation. AL was supported in part by NSF Grant 2307888.
\vspace{10pt}

\appendix

\section{Thick axisymmetric rings}
\label{app:ThickRings}

This appendix explores the relation between the decay rate of the visibility amplitude of a thick axisymmetric ring and its width.

In polar coordinates, the radio visibility $V(u,\varphi)$ of an image $I(\rho,\phi)$---that is, its Fourier transform \eqref{eq:ComplexVisibility}---is
\begin{align}
	\label{eq:PolarVisibility}
	V(u,\varphi)=\int_0^\infty\!\!\int_0^{2\pi}I(\rho,\phi+\varphi)e^{-2\pi iu\rho\cos{\phi}}\rho\ed\rho\ed\phi.
\end{align}
For an axisymmetric image with a purely radial profile $I(r)$, this is simply the zero-order Hankel transform
\begin{align}
	V(u)=H_0[I(\rho)]
	=\int_0^\infty2\pi\rho J_0(2\pi u\rho)I(\rho)\ed\rho,
\end{align}
which is self-inverse ($H_0^2=I$), so $I(\rho)=H_0[V(u)]$.

\subsection{Convolution theorem for Hankel transform}

If two axisymmetric images $I_1(\rho)$ and $I_2(\rho)$ have the visibilities $V_1(u)=H_0[I_1(\rho)]$ and $V_2(u)=H_0[I_2(\rho)]$, then their product image has visibility $V(u)=H_0[I_1(\rho)I_2(\rho)]$ given by
\begin{align}
	\label{eq:ConvolutionTheorem}
	V(u)=\int_0^\infty\!\!\int_0^{2\pi}V_1(U)V_2(u')u'\ed u'\ed\varphi,
\end{align}
with $U^2=u^2+u'^2-2uu'\cos{\varphi}$.
Since the zero-order Hankel transform is self-inverse, this formula also holds with $I\leftrightarrow V$ interchanged.

\subsection{Infinitely thin ring}

An infinitely thin ring of radius $r$ (normalized to have unit total flux) has radial profile
\begin{align}
	I_\delta(\rho)=\frac{1}{2\pi r}\delta(\rho-r),
\end{align}
and corresponding visibility
\begin{align}
	V_\delta(u)=J_0(2\pi ru).
\end{align}

\subsection{General thick axisymmetric ring}

Consider another image with some radial profile $I_w(\rho)$ and associated visibility $V_w(u)$.
By \eqref{eq:ConvolutionTheorem}, the product visibility $V(u)=V_\delta(u)V_w(u)$ corresponds to an image
\begin{align}
	I(\rho)&=\int_0^\infty\!\!\int_0^{2\pi}I_w(R)\frac{\delta(\rho'-r)}{2\pi r}\rho'\ed\rho'\ed\phi\notag\\
	\label{eq:ImageConvolution}
	&=\int_0^{2\pi}\frac{\ed\phi}{2\pi}I_w\pa{\sqrt{\rho^2+r^2-2\rho r\cos{\phi}}},
\end{align}
since $R^2=\rho^2+\rho'^2-2\rho\rho'\cos{\phi}$.
This leads to a simple idea: any bump $I_w(\rho)$ of width $w$ at the origin creates a ring image $I(\rho)$, and vice versa.

\subsection{Example: Gaussian ring}

The Gaussian ring of width $w$, diameter $d$, and unit total flux ($V(0)=1$), has a visibility
\begin{align}
	\label{eq:GaussianVisibility}
	V(u)=J_0(2\pi ru)e^{-2\pi^2w^2u^2}.
\end{align}
This is of the product form $V(u)=V_\delta(u)V_w(u)$ with $V_w(u)=e^{-2\pi^2w^2u^2}$, which corresponds to a unit-flux Gaussian bump of width $w$:
\begin{align}
	\label{eq:GaussianProfile}
	I_w(\rho)=\frac{1}{2\pi\sigma^2}e^{-\frac{\rho^2}{2w^2}}.
\end{align}
By \eqref{eq:ImageConvolution}, the Gaussian ring image has a radial profile
\begin{align}
	I(\rho)&=\frac{1}{(2\pi)^2rw^2}\int_0^\infty\!\!\int_0^{2\pi}e^{-\frac{\rho^2+\rho'^2-2\rho\rho'\cos{\phi}}{2w^2}}\notag\\
    &\qquad\qquad\qquad\qquad\qquad\times\delta(\rho'-r)\rho'\ed\rho'\ed\phi\notag\\
	&=\frac{1}{2\pi rw^2}\int_0^\infty e^{-\frac{\rho^2+\rho'^2}{2w^2}}I_0\pa{\frac{\rho\rho'}{w^2}}\delta(\rho'-r)\rho'\ed\rho'\notag\\
	&=\frac{1}{2\pi w^2}e^{-\frac{d^2}{8	w^2}}I_0\pa{\frac{d\rho}{2w^2}}e^{-\frac{\rho^2}{2w^2}},
\end{align}
where $I_0(x)$ is a modified Bessel function of the first kind.
Its name is justified because as long as the ring diameter is large enough that the intensity is small near the origin, this profile is indistinguishable from a Gaussian of width $w$ and radius $r$:
\begin{align}
	I(\rho)\stackrel{d\gg1}{\approx}\frac{1}{(2\pi)^{3/2}\sqrt{r\rho}w}e^{-\frac{(\rho-r)^2}{2w^2}}.
\end{align}

\subsection{Example: Lorentzian ring}

The Lorentzian (Cauchy distribution) of width $w$ is
\begin{align}
	I_w(\rho)=\frac{1}{\rho^2+w^2}.
\end{align}
Famously, all of its moments diverge.
In particular, this image has infinite flux and its visibility is logarithmically divergent at the origin:
\begin{align}
	V_w(u)=2\pi K_0(2\pi wu)
	\stackrel{u\to0}{\approx}2\pi\log\pa{\frac{1}{2\pi u}},
\end{align}
where $K_0(x)$ is a modified Bessel function of the second kind.
The Lorentzian ring of width $w$ and radius $r$ has visibility
\begin{subequations}
\begin{align}
	V(u)&=2\pi J_0(2\pi ru)K_0(2\pi wu)\\
	&\stackrel{u\to\infty}{\approx}\frac{2\pi^2J_0(2\pi ru)}{\sqrt{wu}}e^{-2\pi wu},
\end{align}
\end{subequations}
which on long baselines decays like a linear exponential.

\subsection{Example: ``Smooth bump'' ring}

The ``smooth bump'' of width $w$ is the function
\begin{align}
	f_w(x)=
	\begin{cases}
		e^{-\frac{w^2}{w^2-x^2}}
		&x\in[-w,w],\\
		0&\text{otherwise}.
	\end{cases}
\end{align}
It defines a (normalizable) radial bump of width $w$
\begin{align}
	I_w(\rho)=cf_w(\rho),\quad
	c^{-1}=2\pi\int_0^\infty f_w(\rho)\rho\ed\rho,
\end{align}
whose convolution with $I_\delta(\rho)$ produces a ring image $I(\rho)$ with compact support localized in a band of width $2w$.
The Fourier transform of $f_1(x)$ behaves asymptotically as \cite{Johnson2015}
\begin{align}
	\tilde{f}_1(k)\stackrel{k\to\infty}{\approx}2\Re\br{\sqrt{\frac{-i\pi}{\sqrt{2i}k^{3/2}}}e^{ik-\frac{1}{4}-\sqrt{2ik}}}.
\end{align}
Therefore, we expect that on long baselines, the visibility $V(u)$ of a ``smooth bump'' ring of unit width will scale as
\begin{align}
	V(u)\stackrel{u\to\infty}{\propto}\frac{J_0(2\pi ru)}{u^{3/4}}e^{-\sqrt{u}},
\end{align} 
which indeed appears to be the case numerically.

\subsection{An observation}

In the three examples above, the visibility of a ring of width $w$ asymptotically behaves as $V(u)\sim e^{-c(wu)^p}$ for some constants $c$ and $p$.
Mathematical properties of the profile can impose stringent restrictions on these constants.
For instance, if the profile is analytic, then $p\ge1$ by the Paley-Wiener theorem, as exemplified by the Gaussian and Lorentzian rings (meanwhile, the smooth bump has $p=0.5$ but is not analytic).
It seems worthwhile to explore the set of possible values of $c$ and $p$ found in phenomenological models and to investigate whether a robust connection to the ring width $w$ can be established.

\section{No universal regime for thick rings}
\label{app:NoUniversalRegime}

In Sec.~\ref{sec:RingSignatures}, we discussed how a relatively thicker ring (such as the $n=1$ photon ring in many models) may not display a universal regime.
In this appendix, we examine explicitly how the universal regime opens or closes up as a function of the width of a Gaussian ring.

A ring of width $w$ and diameter $d$ has one dimensionless parameter: its width-to-diameter ratio, or thickness
\begin{align}
	t=\frac{w}{d}\in\left[0,\tfrac{1}{2}\right),
\end{align}
This thickness cannot be too large: $t\lesssim\frac{1}{2}$ is necessary to have a ring rather than a disk.
In the limit $t\to0$, we always recover the infinitely thin ring $I_\delta(\rho)$ with visibility $V_\delta(u)=J_0(2\pi ru)$.
A generic ring $I(\rho)$ lies in between these two extremes:
\begin{itemize}
	\item A thick ring has $0\ll t\lesssim\frac{1}{2}$.
	\item A thin ring has $0<t\ll\frac{1}{2}$.
\end{itemize}
Its visibility $V(u)$ has two dimensionless scales $U=du$ and $W=wu$.
These are not independent, but related by
\begin{align}
	0<W=tU<U.
\end{align}
It is best to view the visibility as a function $V(U)$ that has a spacing of nulls $\Delta U\approx1$ and exhibits three regimes:
\begin{enumerate}
	\item $W<U<1$, or $u<\frac{1}{d}<\frac{1}{w}$: in this regime, the visibility does not yet resolve the ring, as it has not even reached its first null at $U\sim1$ or $u\sim\frac{1}{d}$.
	\item $W<1<U$, or $\frac{1}{d}<u<\frac{1}{w}$: in this regime, the visibility resolves the ring (at least one null), but not yet its width.
	\item $1<W<U$, or $\frac{1}{d}<\frac{1}{w}<u$: in this regime, the ring has been fully resolved out.
\end{enumerate}
This applies to both thick and thin rings, but for thin rings only, a qualitatively new behavior can emerge in the second regime.
As usual, the reason is that if a system has a small dimensionless parameter, then it ought to exhibit a large separation of scales (and vice versa).

Here, if $0<t\ll\frac{1}{2}$, then it is possible to simultaneously have $1\ll U$ and $tU\ll1$, opening up a new regime $W\ll1\ll U$, or $\frac{1}{d}\ll u\ll\frac{1}{w}$: the smallness of $t=\frac{W}{U}$ is equivalent to the large separation of scales needed to open up this new regime.

This is the universal regime.
It is universal because in it we may approximate $W\approx0$ and hence set $t\approx0$ while still keeping $U\gg1$, which means that it is possible to stay in this regime while forgetting about the width---and radial profile---of the ring.
Therefore, the visibility of any thin ring in this regime must tend to $V_\delta(u)$.

\subsection{Example: Gaussian ring}

We now explore this in the context of a Gaussian ring of width $w$ and diameter $d=2r$, with visibility \eqref{eq:GaussianVisibility} or
\begin{align}
	\label{eq:GaussianProfile}
	V(U)=J_0(\pi U)e^{-2\pi^2W^2}
	=J_0(\pi U)e^{-2t^2\pi^2U^2}.
\end{align}

In regime 1, both $U$ and $W=tU$ are small, so we may expand in $U\ll1$ to find the second-order approximation
\begin{align}
	V(U)\approx1&-\pa{\frac{1}{4}+2t^2}\pi^2U^2\notag\\
	\label{eq:GaussianRegime1}
    &+\pa{\frac{1}{64}+\frac{t^2}{2}+2t^4}\pi^2U^2,
\end{align}
which we expect to be valid for $U<1$ and any $t$.

\begin{figure*}[ht!]
	\includegraphics[width=0.7\columnwidth]{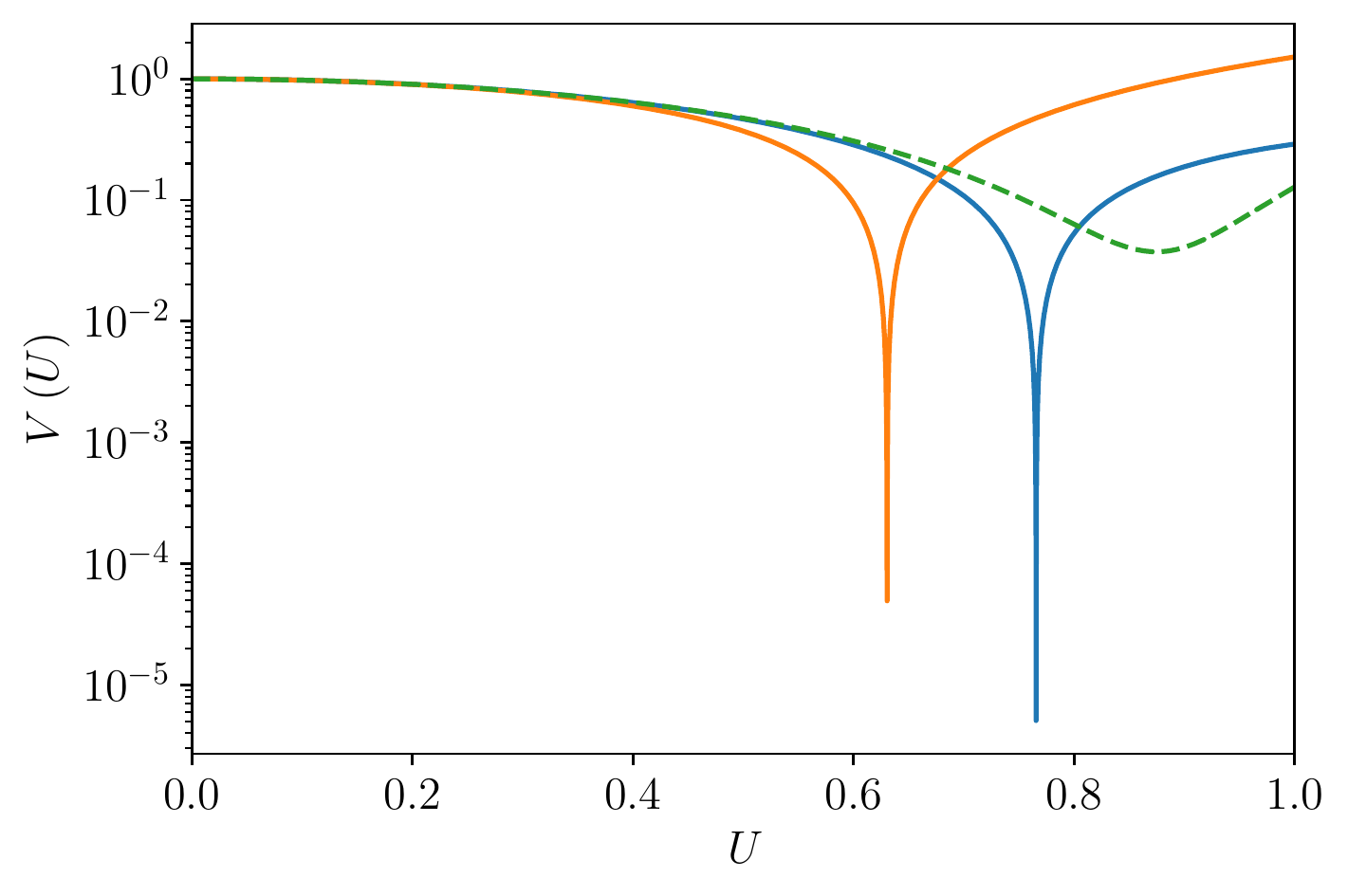}
	\includegraphics[width=0.7\columnwidth]{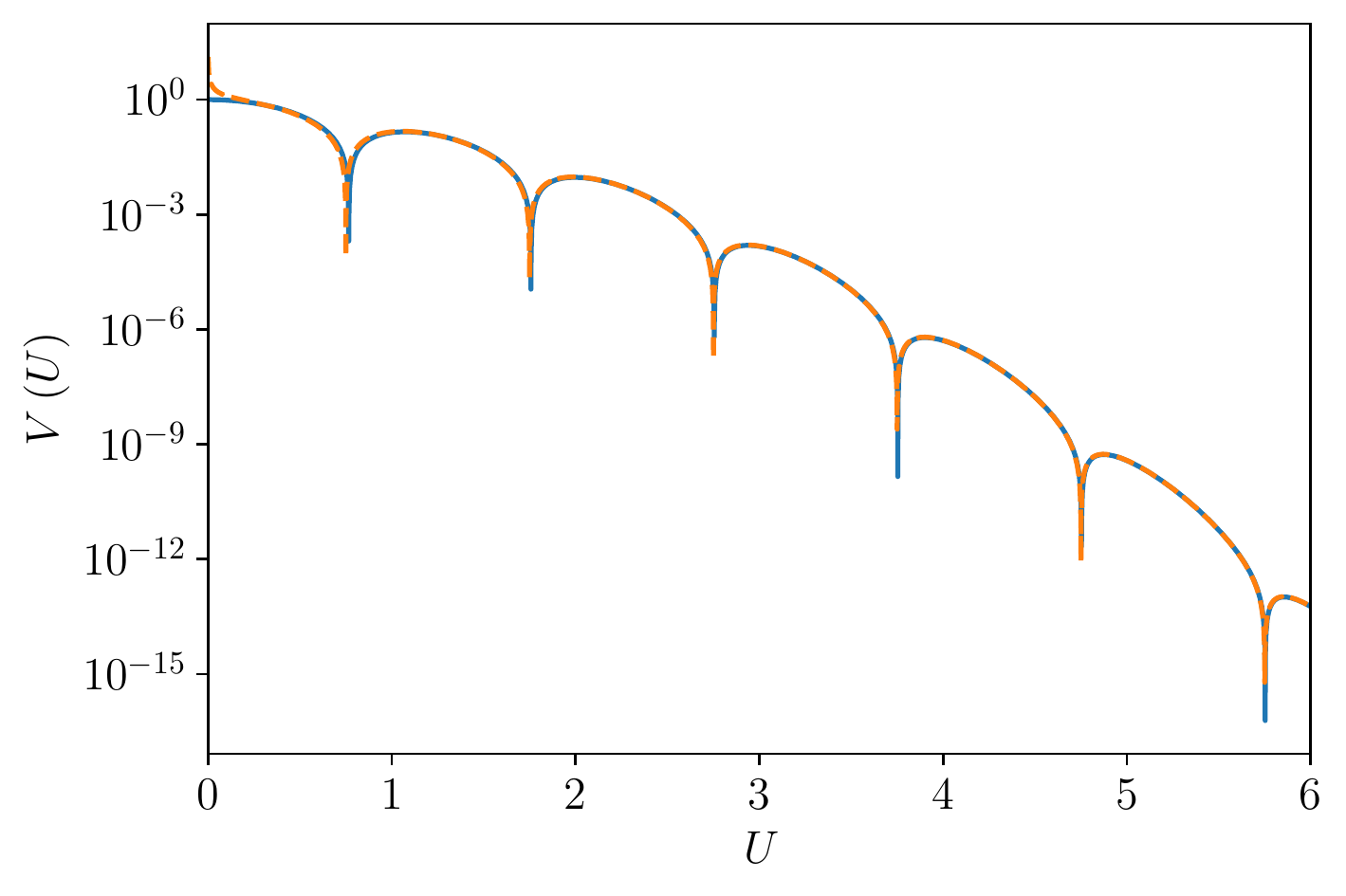}\\
	\includegraphics[width=0.7\columnwidth]{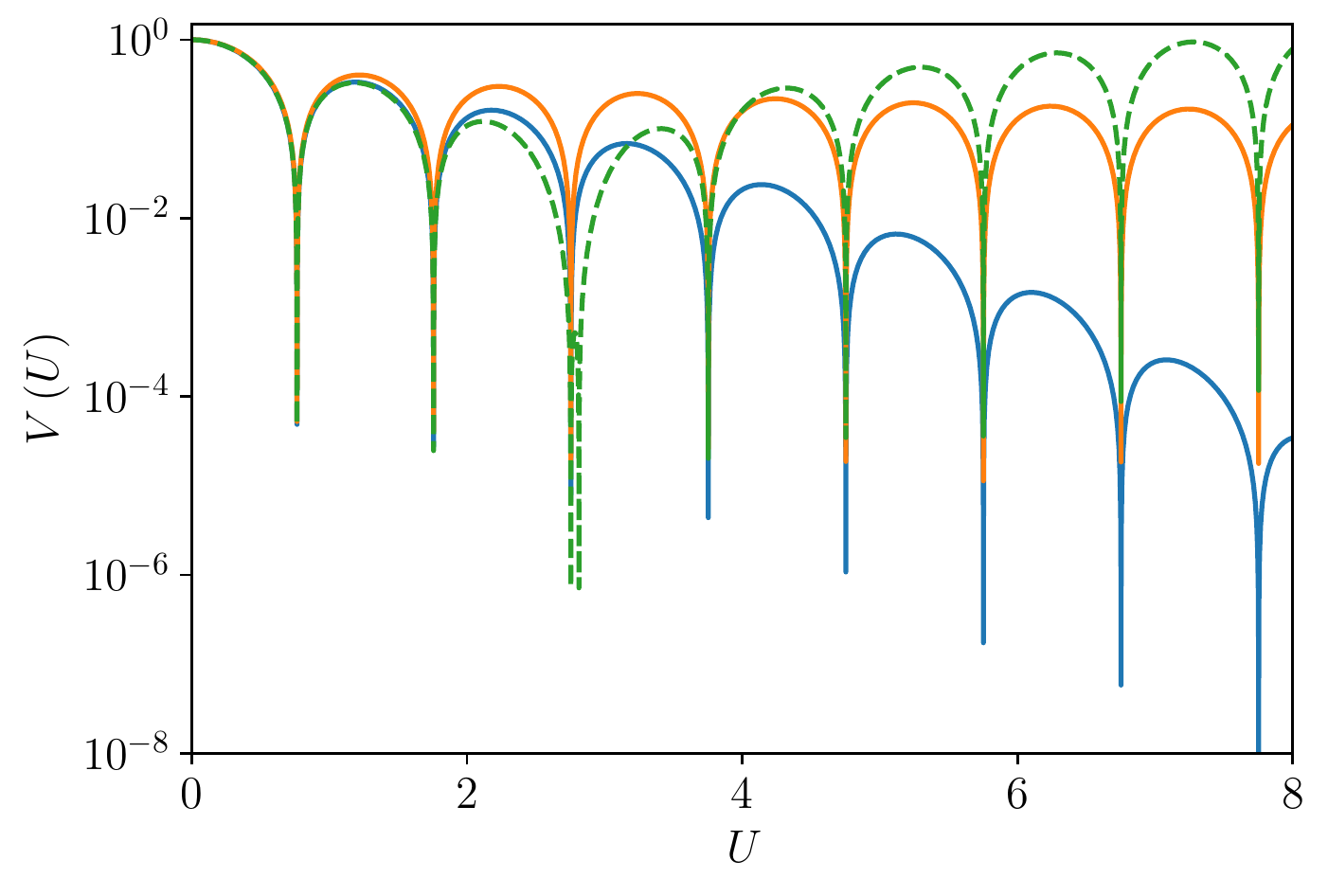}
	\includegraphics[width=0.7\columnwidth]{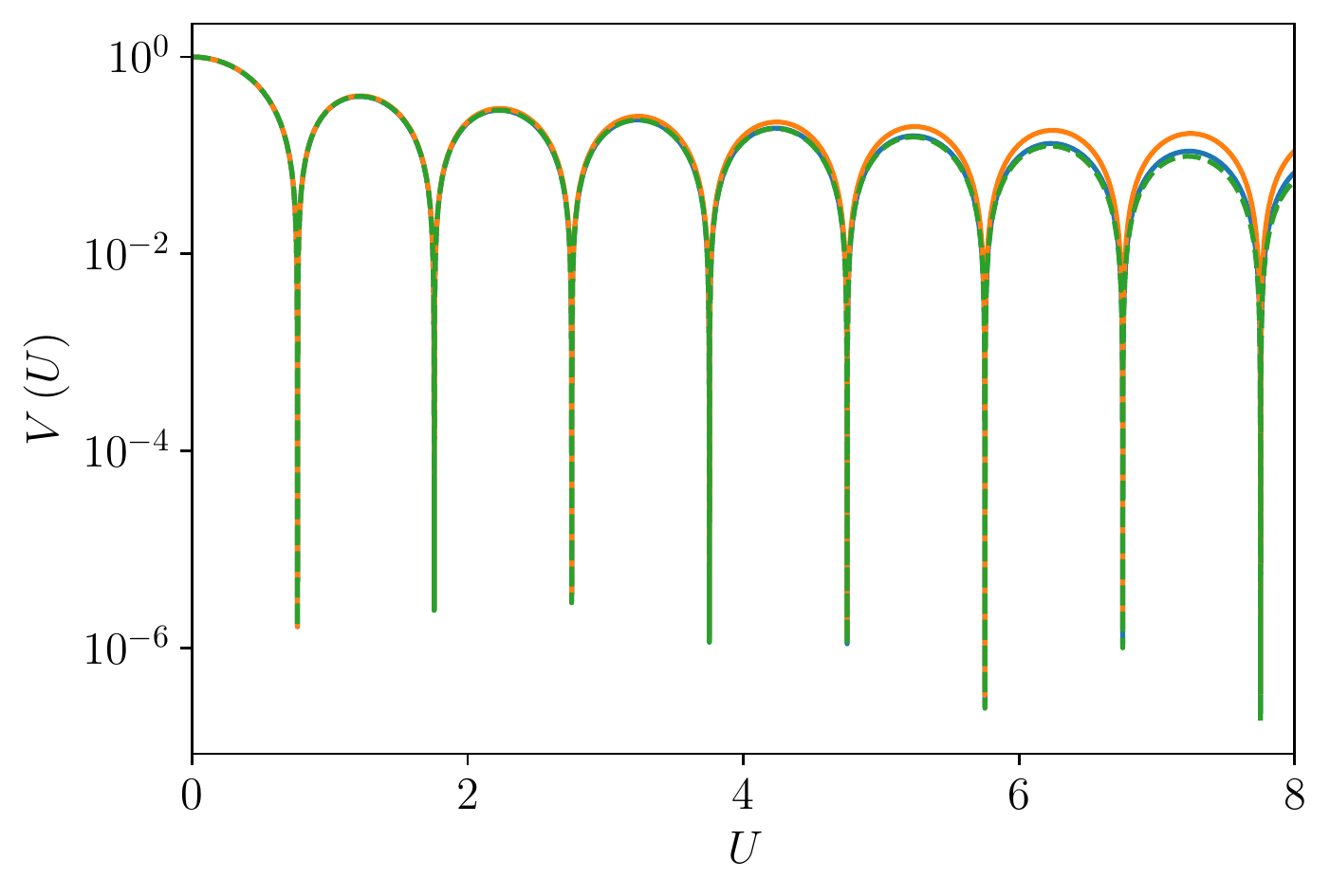}
	\caption{\textbf{Top left:} Exact visibility amplitude $|V(U)|$ (blue) of a Gaussian ring with $t=5\%$ against its leading (orange) and subleading (green) approximations \eqref{eq:GaussianRegime1} in regime 1, showing excellent agreement in the expected range $U\ll1$.
	\textbf{Top right:} Exact visibility amplitude $|V(U)|$ (blue) of a Gaussian ring with $t=20\%$ against its leading approximation \eqref{eq:GaussianRegime3} in regime 3, showing excellent agreement in the expected range $U\gg1$.
    \textbf{Bottom left:} Exact visibility amplitude $|V(U)|$ (blue) of a relatively thicker Gaussian ring with $t=8\%$ against its leading (orange) and subleading (green) approximations \eqref{eq:GaussianRegime2} in regime 2.
    \textbf{Bottom right:} Same for a narrower ring of thickness $t=2\%$.
	In both cases, we find excellent agreement in the expected range $1\ll U\ll(2\pi t)^{-1}$, where the profile follows the universal form \eqref{eq:UniversalForm} of an infinitely thin ring.}
	\label{fig:GaussianRing}
\end{figure*}

In Fig.~\ref{fig:GaussianRing}, we confirm this by plotting the exact $|V(U)|$ (blue) against its leading (orange) and subleading (green) approximations for $t=5\%$, finding good agreement in the expected range $U\ll1$.

In regime 3, both $U$ and $W=tU$ are large, so we may expand in $U\gg1$ to obtain the leading approximation
\begin{align}
	\label{eq:GaussianRegime3}
	V(U)\approx\frac{\cos{\pi U}+\sin{\pi U}}{\pi\sqrt{U}}e^{-2t^2\pi^2U^2},
\end{align}
which we plot in Fig.~\ref{fig:GaussianRing} (orange) against the exact $|V(u)|$ (blue) for a ring of width $t=20\%$, again finding excellent agreement in the expected regime $U\gg1$.

We note that the approximations obtained in these regimes are not in any sense universal: in particular, they depend on the thickness $t$ of the ring and would differ for a non-Gaussian radial profile.

Finally, in regime 2, $U$ is large but $W=tU$ is small, so we cannot expand in $U$.
We have no recourse but to expand in $t\ll1$ to find
\begin{align}
	\label{eq:GaussianRegime2}
	V(U)\approx J_0(\pi U)\br{1-2t^2\pi^2U^2+\ldots},
\end{align}
which is only supposed to be a valid approximation for
\begin{align}
	\label{eq:UniversalRange}
	1<U\lesssim\frac{1}{2\pi t}.
\end{align}
Thus we only expect this approximation to be good as $t\to0$, in which case it holds over a very large range and forgets about the width (radial profile) of the ring to take the universal form
\begin{align}
	\label{eq:UniversalForm}
	V(U)\approx J_0(\pi U)=V_\delta(U).
\end{align}
In Fig.~\ref{fig:GaussianRing}, we confirm this by plotting the exact visibility $|V(u)|$ (blue) against its leading (orange) and subleading (green) approximations \eqref{eq:GaussianRegime2}, first for a relatively thicker ring with $t=8\%$ and next for a thinner ring with $t=2\%$.

Remarkably, we find that the agreement is good up to $U\lesssim2$ in the first case, and up to $U\lesssim8$ in the second case, exactly as predicted by \eqref{eq:UniversalRange}.
At the lower end, we find that for the Gaussian ring, the universal formula applies all the way down to $U=0$, but this is merely a coincidence: this would not be true of the Lorentzian ring, for instance, whose visibility diverges at the origin.

\bibliographystyle{apsrev4-1}
\bibliography{n1}

\end{document}